\documentclass[lettersize,journal]{IEEEtran}
\usepackage[dvips]{graphicx}
\usepackage{amsmath,amsfonts}
\usepackage{algorithmic}
\usepackage{algorithm}
\usepackage{array}
\usepackage{mathtools}
\usepackage[caption=false,font=normalsize,labelfont=sf,textfont=sf]{subfig}
\usepackage{caption}
\usepackage{textcomp}
\usepackage{stfloats}
\usepackage{url}
\usepackage{verbatim}
\usepackage{graphicx}
\usepackage{cite}
\usepackage{xcolor}
\usepackage{amssymb}
\usepackage{tabularx}
\usepackage{array}
\usepackage{todonotes}
\usepackage[greek,english]{babel}
\newcolumntype{Y}{>{\raggedright\arraybackslash}X}
\allowdisplaybreaks

\usepackage{tikz}

\newcommand{\bluecircle}[1]{%
  \tikz[baseline=(X.base)]{
    \node (X) [circle, fill=cyan!30, text=black, inner sep=0.5pt] {$#1$};
  }%
}
\newcommand{\redcircle}[1]{%
  \tikz[baseline=(X.base)]{
    \node (X) [circle, fill=red!30, text=black, inner sep=0.2pt] {$#1$};
  }%
}
\usepackage{fontspec}
\usepackage{hyperref} 

\begin{document}

\title{A Copper-Inducible Yeast Biosensor with Bioluminescent Readout: Mathematical Modeling, Experimental Validation, and Photodetector Coupling for Molecular Communication in IoBNT}

\author{Nikolaos~Ntetsikas,~\IEEEmembership{Student Member,~IEEE,} 
        Styliana~Kyriakoudi, Ariel~Klavaris,
        Antonis~Kirmizis,
        Chrysostomos Chrysostomou,~\IEEEmembership{Member,~IEEE,}
        Ian F. Akyildiz,~\IEEEmembership{Life Fellow,~IEEE} and
        Marios Lestas,~\IEEEmembership{Member,~IEEE} 
\thanks{N. Ntetsikas and C. Chrysostomou are with Frederick Research Center and the School of Engineering, Frederick University, Nicosia, 1036, Cyprus (email: st023422@stud.frederick.ac.cy, ch.chrysostomou@frederick.ac.cy).}
\thanks{S. Kyriakoudi is with the Laboratory of Epigenetics and Gene Regulation, Department of Biological Sciences, University of Cyprus, Nicosia, 2109, Cyprus and Frederick Research Center, Nicosia, 1036, Cyprus (email: res.ks@frederick.ac.cy).}
\thanks{A. Klavaris and A. Kirmizis are with the Laboratory of Epigenetics and Gene Regulation, Department of Biological Sciences, University of Cyprus, Nicosia, 2109, Cyprus (email: klavaris.ariel@ucy.ac.cy, kirmizis.antonis@ucy.ac.cy).}
\thanks{M. Lestas is with the Department of Electrical Engineering, and Computer Science and Engineering,  Cyprus University of Technology, Limassol, 3036, Cyprus (email: marios.lestas@cut.ac.cy).}
\thanks{Ian F. Akyildiz is with Truva Inc., Alpharetta, Georgia, 30022, United States of America (email: ian@ianakyildiz.com).}}



\maketitle

\begin{abstract}
Engineered living biosensors constitute promising model organisms for bringing Molecular Communications (MC) closer to practical applications. However, challenges associated with their development and characterization have resulted in the relevant literature being limited in the MC domain, with the research area remaining largely unexplored despite its prospects. In this paper, we contribute in this direction by leveraging engineered yeast as a basis to detect heavy metals such as copper. The biosensor's development is supported by a validated mathematical characterization and analysis for possible integration in an MC system. Different from the existing literature, mathematical representations of the major biological processes within the yeast copper detoxification pathway are integrated into a comprehensive framework that predicts the time evolution of the $CUP1$ gene, which is treated as the system output. In the engineered circuit, \textit{CUP1} was replaced with \textit{GFP}, therefore making GFP fluorescence the final system output. Εxperimental analysis including mRNA transcript quantification and fluorometric measurements, validate the GFP reporting model to a satisfactory degree. This establishes the proposed model as a useful tool for virtual experimentation and the developed testbed as a suitable platform for further engineering of yeast-based copper sensors. The theoretical analysis is further extended by incorporating a self-sustained bioluminescent reporter for yeast, replacing GFP and enabling optical output without external light excitation. The optical response is coupled to a photodetector model, rendering the system suitable for integration into mobile monitoring devices such as Smart Pills. Finally, sensitivity analysis reveals changes in amplitude, peak time, and pulse width, resulting from variation of key model parameters and chemical species. The predicted photocurrent profile is found to be within the femto-Ampere (fA) range and exceeds the estimated dark current, suggesting that detection may be feasible under idealized conditions. The presented characterization of the yeast receiver as a copper biosensor, constitutes the first step towards the development and characterization of a complete end-to-end MC based system for gut theranostics after coupling with suitable engineered probiotic senders.
\end{abstract}





 \maketitle

\section{Introduction}\label{intro}
Over the past decade, MC has attracted increasing attention in the scientific community, driven by advances in both theory and experimentation. 
To date, the literature has predominantly focused on theoretical foundations, including computational frameworks and modeling techniques for channel characterization, localization, and information rate optimization. In parallel, MC has been positioned as an enabling layer for applications ranging from industrial scenarios such as IoT-based inspection and monitoring in pipelines \cite{AlZubi2025} to healthcare, including the Internet of Bio-Nano Things (IoBNT) for disease monitoring and treatment \cite{Bulasara}, intra-vessel monitoring \cite{Felicetti2025}, and particulate targeted drug delivery for tumors \cite{Kamal2025}.

Nevertheless, the translation of the solid theoretical base into realistic settings requires experimental testbeds that can reproduce key propagation and signal transduction mechanisms at relevant scales. Consequently, a growing body of work has developed MC testbeds at both the macroscale and microscale. Microscale implementations frequently incorporate biological entities as transmitters and receivers coupled through a fluidic channel, thereby enabling molecular propagation in controlled environments. However, among the reported biological testbeds, prokaryotic organisms are most commonly adopted, whereas only a limited number of studies consider eukaryotic communication units \cite{Felicetti2014,Nakano2008,Ntetsikas2024,Ntetsikas2025}. Importantly, irrespective of organism class, microorganisms can exhibit measurable responses to external molecular stimuli. Such an attribute, naturally connects MC testbeds to biosensing applications and in particular with living biosensors.   


Bacteria and yeast constitute notable model organisms which have been leveraged for living biosensor development, however, their integration into the MC framework remains limited with a notable bacteria based example reported in \cite{Anderl2024}. The co-integration of living sensing components with embedded electronics for in vivo biomarker monitoring (with the most notable enabling technology being Smart Pills \cite{IndaWebb2023}) underscores the practical relevance of biosensing-based receivers and motivates the potential of MC integration. Moreover, living biosensors can also be envisioned as MC senders e.g. in the context of Organ-on-Chip (OoC) integration \cite{Ling2026}, further strengthening their potential to serve the MC domain.   

Yeast-based biosensors offer advantages over bacterial systems due to their closer cellular and molecular similarity to higher eukaryotes. In particular, Saccharomyces cerevisiae (\textit{S. cerevisiae} hereafter) shares key features with mammalian cells, making it a relevant model for sensing human-related biomarkers. Combined with advances in synthetic biology and genome editing, yeast can be readily engineered to sense targets either through its native pathways or by incorporating heterologous genes. The unique properties of yeast underscore its biosensing potential, particularly for detecting short-lived disease biomarkers in humans \cite{Ntetsikas2024}. However, the capability of MC integrated into yeast biosensors has so far remained largely untapped, therefore limiting their ability to play a pivotal role in human healthcare. 

Previous work on a yeast-based MC testbed \cite{Ntetsikas2024,Ntetsikas2025} developed a mathematical framework for pheromone-mediated yeast cell-to-cell (cell2cell) communication, and was supported by extensive experimental trials to validate theoretical findings. Subsequent research focused on noise analysis tailored for eukaryotic systems, led to the development of Maximum-Likelihood schemes for signal detection. In this work, departing from previous biosensor designs within the MC literature, a yeast biosensor that incorporates the unicellular \textit{S. cerevisiae} as the sensing unit for the detection of cytotoxic copper accumulation in the gut is developed. A comprehensive mathematical model describing the major biological processes and mechanistic interactions involved in the copper detoxification pathway is developed. Particular emphasis is placed on the dynamics of the underlying biochemical reactions, and the interplay among the involved chemical compounds. In response to excessive copper accumulation in their vicinity, \textit{S. cerevisiae} cells are induced to express a detoxifying gene, \textit{CUP1}, which is treated as the system output as well as an indication of potentially toxic heavy-metal levels. Building upon the yeast MC testbed reported in \cite{Ntetsikas2024}, a new testbed is developed and is utilized as a validation tool for the mathematical analysis. Moreover, the theoretical characterization of the biosensor is developed with Smart Pill integration as a guiding objective. Overall, the envisioned MC system includes the application of living senders that can detect and transmit molecular signals in areas such as the gut, and encapsulated MC receivers that are able to forward the detected information to external monitor devices, thereby aiming to advance the IoBNT toward intra-body diagnostics and therapeutics. The contributions of this paper can be summarized as follows:
\begin{itemize}
    \item Dynamic integration of the major reactions involved in the copper detoxification pathway into a system of coupled nonlinear Ordinary Differential Equations (ODEs). To the best of the authors’ knowledge, copper uptake, intracellular dissemination through the metallochaperone pathways, copper-dependent activation of transcriptional regulators, and gene expression have not previously been jointly represented within a dynamic receiver model.
    \item Development of the yeast module of the biosensor as an MC receiver using the new testbed developed. Experimental trials via fluorometric measurements characterize the yeast's response, in both mRNA and protein level. mRNA and fluorescence measurements are also employed to assess the specificity of the proposed biosensor against another chemically related metal, namely zinc.
    \item Cross matching between experimental results and theoretical findings for the GFP reporter, reveals satisfying concordance between the two across multiple concentrations. This renders the mathematical tool presented here as a solid base for further research.
    \item Extension of the receiver model to include bioluminescence as an appropriate reporting output, coupled to a photodiode response. Toward this goal, the GFP reporter models are replaced by Lux-based models. The resulting Lux response is treated as a predictive extension that is not experimentally validated in the present work.
\end{itemize}

The modeling effort here has been centered around the theoretical characterization of the yeast biosensor as a copper receiver, where a direct copper input is assumed. The receiver model developed in this work however, may provide the basis for its integration with a transmission entity and a channel sub-system into a complete MC framework. Communication performance metrics such as achievable data rates and error analysis will also be able to be investigated.

The rest of the paper is organized as follows: In Section~\ref{review} the current state of the art in biosensors is presented. In Section~\ref{description} the overall system under consideration is described, and the design principles of the biosensor are outlined.  The mathematical characterization of the yeast MC receiver followed by integration to a photodetector is discussed in detail in Section~\ref{system model}. In Section~\ref{testbed}, the experimental platform is described, and the methods and materials used for the experiments are provided. In Section~\ref{results} both theoretical and experimental results are presented, and an attempt to validate the GFP predictions is reported. Moreover, the models derived for the extension to a luminescent response, are used to characterize the biosensor's output and conduct a sensitivity analysis to investigate the robustness of the mathematical approach. Finally, Section~\ref{Conclusion} summarizes the paper's findings and provides some directions for future work.

\section{Reported biosensors}\label{review}

A yeast-derived biosensor for copper monitoring in the gut is presented, with Smart Pill integration considered as a target deployment scenario. Prominent Smart Pill biosensors suitable for in vivo operation have been reported in \cite{IndaWebb2023} and \cite{Xu2025}. In \cite{IndaWebb2023}, probiotic bacteria were engineered to detect biomarkers associated with Inflammatory Bowel Disease (IBD), including nitric oxide, hydrogen peroxide (H$_2$O$_2$), tetrathionate, and thiosulfate. In vivo validation was performed in both mice and pigs, and GFP readouts indicated that adequate sensitivity and specificity were achieved for the targeted analytes. The calibrated bacterial sensors were subsequently encapsulated, and GFP was replaced with the \textit{luxABCDE} operon to enable autonomous bioluminescent reporting. With the exception of the thiosulfate sensor, specificity and sensitivity were largely preserved after encapsulation. Analyte information was successfully transmitted to external units, thereby reducing the need for direct sampling from the gastrointestinal tract or from stool.
In \cite{Xu2025}, swallowable bacterial sensors designed to respond to heme within the intestinal environment were reported. Magnetic particles were incorporated to facilitate retrieval from stool, while sodium alginate droplets were used for encapsulation to enhance survival through digestion. Performance was assessed in a colitis mouse model, where microsphere recovery and analysis were completed in approximately 25 minutes. The reported response scaled with disease severity.

Additional pill-form biosensors have been reported in \cite{DelaPaz2022,Zhuang2025,Min2025,Abdigazy2024}. In \cite{DelaPaz2022}, a glucose-monitoring capsule was reported in which two electrodes (forming a glucose biofuel cell (BFC)), nano-circuitry, and a magnetic human-body communication (mHBC) antenna operating from 40 to 200 MHz were integrated for in-gut sensing. Energy harvesting was enabled through glucose-driven BFC operation, thereby reducing reliance on battery sources. The electronic components were insulated using silicone/polyurethane coatings, while an enteric coating was employed to protect the electrodes from the stomach environment. Signal transmission was achieved through voltage-to-frequency conversion, whereby glucose concentration was mapped to the carrier frequency. Sensor calibration was performed in vitro, whereas in situ validation in pigs demonstrated frequency variations in response to different glucose concentrations through supplementation of saline samples. Specificity was assessed by applying alternative chemical inputs, for which baseline frequency operation was maintained. In \cite{Zhuang2025}, a capsule biosensor targeting Reactive Oxygen Species (ROS) associated with IBD was presented, where a blue dye readout was released upon ROS exposure. A modified dextran coating was employed and was dissolved in the presence of ROS, enabling dye release and subsequent stool-based readout. Evaluation in a colitis mouse model yielded 72\% sensitivity and 78\% specificity. In \cite{Min2025}, a multi-analyte capsule platform was reported. It comprised an array of sensing modalities including aptamer-based voltammetric serotonin sensors, enzymatic amperometric glucose sensors, ion-selective potentiometric pH sensors, and impedimetric ionic-strength sensors. In vivo experiments were conducted in mice and rabbits, where fasting protocols were applied prior to administration of dietary samples to trigger secretion of the targeted compounds. In \cite{Abdigazy2024}, a gas-sensing capsule was reported in which an oxygen-responsive membrane was utilized as the sensing element. Power was supplied using coin-cell batteries, while fluorescence emission was generated through membrane-oxygen interaction. Optical excitation was provided by integrated LEDs, and the resulting signal was transmitted via Bluetooth to an external device. Capsule localization was achieved using an external planar wearable coil, through magnetic-field excitation of an embedded magnetic sensor. Validation was performed in vitro using GI tract fluid simulants, while promising results were additionally reported in tests conducted in cow intestines.

Recruitment of biological entities in place of purely artificial sensing components has been considered advantageous, as improved biocompatibility, smoother long-term operation, and reduced health risk due to fewer synthetic materials may be achieved. In addition, enhanced energy efficiency may be enabled, since substantially lower power requirements are typically imposed by biological transduction compared to fully electronic sensing chains. Although living entities in the form of bacteria have been investigated in \cite{IndaWebb2023} and \cite{Xu2025}, MC techniques have not been explicitly integrated to characterize and optimize the communication aspects of the reported systems. Receiver-oriented modeling and analysis have, however, been pursued in a number of bacteria-based MC biosensing studies. In \cite{Amerizadeh2021}, bacteria were engineered to respond to L-rhamnose and to produce GFP as an output. The underlying signal-transduction pathway was explicitly modeled and fitted to experimental measurements. A peak response was reported in less than one hour, therefore demonstrating the feasibility of time-resolved biosensing with engineered cells. In \cite{Terrell2021}, a BioLAN network was reported in which an electronically encoded input was transduced into hydrogen peroxide via electrochemically controlled oxygen reduction. This biomolecular input was then used to trigger bacteria that produced acyl homoserine lactone (AHL), which subsequently diffused toward two additional bacterial populations. In one population, $\beta$-galactosidase production was induced and was converted into an electronic readout through electrochemical oxidation and external detection. In the other population a red fluorescent protein was expressed and was used as a transmission confirmation signal. In \cite{Anderl2024}, a bacteria biosensor was analyzed from an MC perspective, where \textit{S. mutans} was considered to respond to the XIP peptide through an innate signal-transduction pathway, thereby indicating its presence. The luminescent response was simulated and was validated using experimental data. Steady-state and dynamic sensitivity analyses were performed, through which a tradeoff between increased sensitivity and output-response strength was revealed under parameter variation. In \cite{salvadori2024}, \textit{Streptococcus pneumoniae} was proposed as a sensing entity for quorum-sensing molecules that promote pathogenic colonization and can lead to severe host complications. The bacteria were engineered to respond to AI-2 and CSP molecules. Luminescent reporting was enabled through expression of the firefly luciferase gene.

In contrast to prokaryotic biosensors, a broad range of yeast-based biosensors has been reported in the literature, spanning diverse applications and introducing distinct challenges and opportunities \cite{Wahid2022}. In \cite{Bisquert2024}, a whole-cell yeast biosensor was engineered for melatonin detection in a fluidic medium through modification of native G Protein Coupled Receptors (GPCR). Evaluation was performed in growth medium supplemented with melatonin as well as in wine samples. There, a tradeoff between detection range and signal-to-noise ratio was identified and specificity was demonstrated. A related GPCR-based design was reported in \cite{Lind2024}, where the yeast mating pathway was engineered to express colorimetric reporters (lycopene, proviolacein, and prodeoxyviolacein). Based on experimental trials under $\alpha$-factor stimulation, lycopene was found to provide the most robust response, thereby suggesting a practical alternative to fluorescent and luminescent reporters. Colorimetric reporting through lycopene was also employed in \cite{Ostrov2017}. On-site pathogen surveillance was enabled by introducing a lycopene production pathway into \textit{S. cerevisiae} and embedding the engineered cells in a dipstick format suitable for direct contact with the liquid under test. Stool-based biosensing was investigated in \cite{Dacquay2024}, where doxycycline presence was reported using \textit{Saccharomyces boulardii}. Specifically, the \textit{ADE2} locus was positioned between two Cre recognition sites, such that Cre expression in response to doxycycline resulted in a visible pigmentation switch from white to red. Successful recovery of red-pigmented yeast from stool was demonstrated in mouse trials. Environmental sensing has also been reported; in \cite{Mendes2024}, \textit{S. cerevisiae} was engineered to produce luciferase in response to the fungicide tebuconazole in water. Yeast sensing platforms targeting metal detection have additionally been demonstrated. In \cite{Wahid2025}, a prototype capable of sensing a range of copper concentrations was presented, where engineered \textit{S. cerevisiae} cells were interfaced with a screen-printed electrode in an electrolyte medium. Copper-induced disruption of glucose metabolism was translated into a reduction of intracellular current density, which was captured by the electrodes and transmitted via Bluetooth to an external unit for processing and visualization. A field-portable and storable \textit{S. cerevisiae} biosensor that has progressed to a commercial product was described in \cite{Lobsiger2019}, where detection of endocrine-disrupting chemicals was enabled (YestroSens). 

It is noted that the efforts in \cite{Wahid2025} and \cite{Lobsiger2019} are distinguished by the inclusion of real-time transmission of detection events to external units, which constitutes a key requirement for practical yeast biosensing and aligns with IoBNT-oriented operation. Nevertheless, the yeast biosensors summarized above have largely been developed as laboratory or application-driven prototypes, with the mathematical characterization that can support subsequent analysis and integration in an MC context largely missing. This work overcomes these limitations, by offering a flexible practical implementation platform for MC receiver/sender development, as well the assorted mathematical characterization which can base MC-type analysis, design and implementation. The flexibility stems from the genetic amenability of yeast which can support a wide range of biomarker detection, as well as multiple modes of integration in existing systems. This is crucial in the realization of MC use cases in practical applications.

\section{System Description} \label{description}
The system under consideration is based on the copper detoxification pathway of \textit{S. cerevisiae} cells \cite{Shi2020}. Copper, which is naturally encountered across biological environments, is essential for multiple cellular processes, serving as a co-factor in enzymatic reactions and contributing to the structural stabilization of various proteins. However, excessive copper accumulation and unregulated uptake can impair protein function and promote the generation of ROS, thereby leading to cellular toxicity. Owing to this dual role, mechanisms for maintaining intracellular copper homeostasis have evolved across life forms, including regulated copper transport, chelation, and efflux.

In the extracellular environment, copper ions are predominantly found in the cupric form. Upon reaching the vicinity of \textit{S. cerevisiae} cells, cupric ions are reduced to cuprous ions by the yeast reductases Fre1 and Fre2. Following internalization, copper ions are distributed via metallochaperones, through which copper is delivered to specific copper-binding proteins within the cell. Under elevated copper conditions, the copper-detoxification pathway culminates in the upregulation of genes that preserve homeostasis, primarily by limiting further uptake and enabling copper sequestration, thereby preventing toxicity. Detoxification is mainly mediated by the metallothionein \textit{CUP1}, through which copper is sequestered and the pool of free intracellular copper ions is reduced. 

\begin{figure}[htpb]
  \centering
  \subfloat[WT \textit{S. cerevisiae} cell responding naturally to copper.]{\includegraphics[width=2.5in]{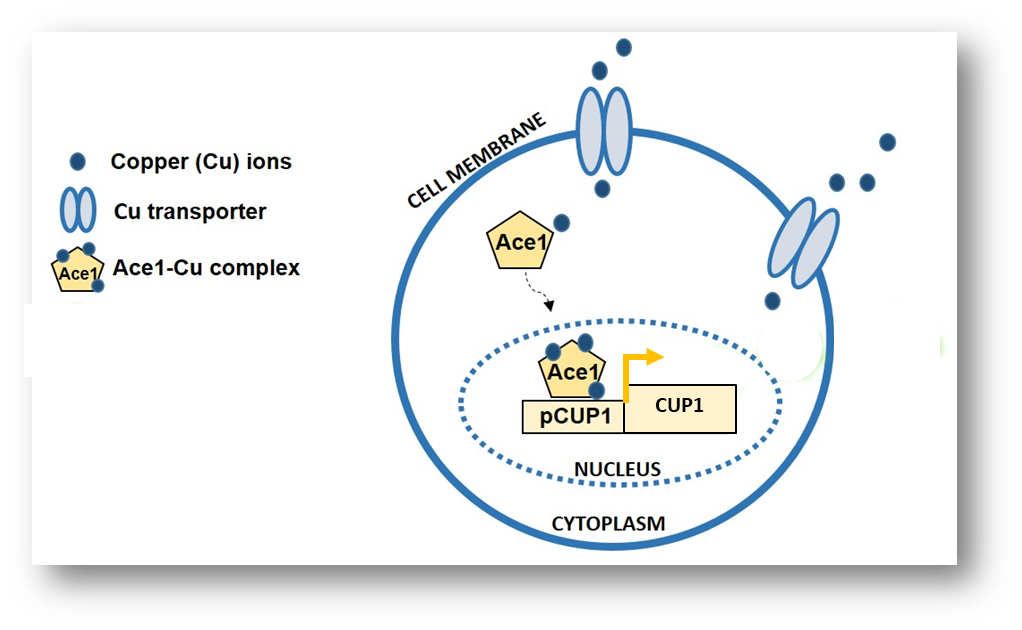}%
 \label{cup1 yeast}}
  \hfill
  \subfloat[\textit{S. cerevisiae} MC receiver for fluorescent biosensing.]{\includegraphics[width=2.5in]{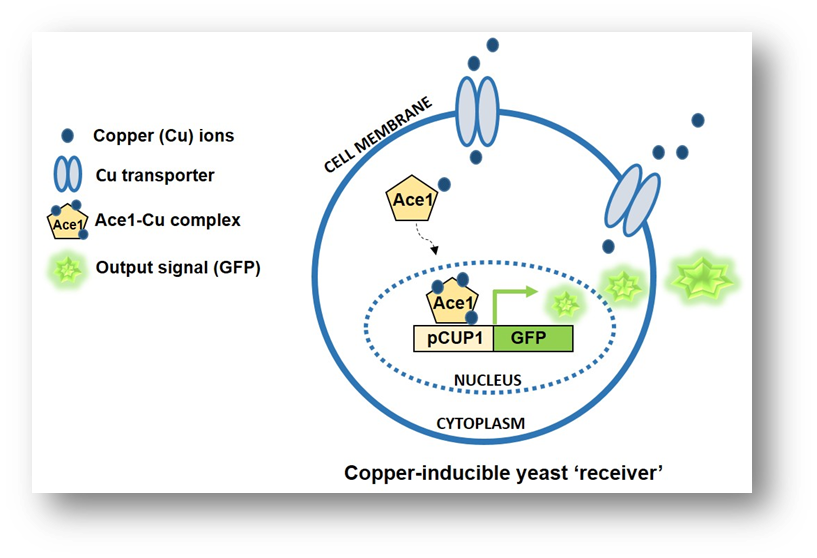}%
 \label{gfp yeast}}
  \caption{Natural response of a yeast cell to copper (figure above) and genetically modified response for copper sensing with GFP as the output (figure below).}
  \label{cells}
\end{figure}

The system considered in this work comprises a whole-cell yeast copper receiver, as illustrated in Fig.~\ref{cup1 yeast}, within a biosensing scenario in which reporting elevated copper levels can serve as an indicative tool. The yeast testbed developed in this work is utilized for direct Cu stimulation of yeast cells, thereby enabling quantification of the detoxification response in wild-type (WT) cells through replacement of the endogenous \textit{CUP1} gene with a reporter protein. The theoretical analysis reported herein follows the same framework, where a direct copper stimulus is assumed by the mathematical model and GFP dynamics are computed for validation purposes. 

Although GFP reporting provides a straightforward means for quantifying gene expression and, consequently, for integrating MC principles into the overall system, fluorescence measurements typically require external excitation light sources. In mobile MC applications where miniaturization is critical, such as Smart Pills, reliance on additional optical components is preferably minimized, while self-sustained operation is desired. To this end, a self-contained bioluminescent reporting scheme is considered here as a replacement for the GFP reporter employed previously. This enables direct coupling of the yeast optical output to the photodetector response.

Bioluminescence in living cells is generated by luciferase systems encoded in the organism’s genome, where expression can be constitutive or induced under specific stimuli. In such systems, light emission is produced through an enzymatic reaction between a luciferase enzyme and one or more chemical substrates. Although Lux systems are naturally present in several bacterial species, this capability is absent in wild-type yeast. This fact motivates genetic engineering when self-sustained optical reporting is sought for MC-integrated biosensing applications.

Among the luciferase systems considered for yeast, integration of the NanoLuc luciferase gene (\textit{yNluc}) and of the bacterial Lux cassette has been investigated in \cite{Masser2016} and \cite{Gupta2003}, respectively. The approach in \cite{Masser2016} is characterized by relatively simple biochemical processes and straightforward integration methods; however, external supplementation of the substrate (furimazine) is required, which would hinder truly standalone operation if such a reporter was adopted in a yeast-derived biosensor.

In contrast, although a higher level of complexity is involved, the Lux system described in \cite{Gupta2003} is self-sufficient, as the required substrates can be produced internally and recycled. In that work, expression of the \textit{luxABCDE} cassette from \textit{Photorhabdus luminescens} was implemented, comprising the enzymes LuxA, LuxB, LuxC, LuxD, and LuxE, while co-expression of the flavin oxidoreductase gene \textit{FRP} from \textit{Vibrio harveyi} was additionally employed. Collectively, these enzymes enable the bioluminescent pathway described in \cite{TINIKUL2020}, relying solely on the cell's resources; however, several steps of the overall process are still not fully understood. The genes \textit{luxA} and \textit{luxB}, which encode the luciferase subunits, are commonly treated as a functional dimer and are referred to as \textit{luxAB} in the literature.

\begin{figure}[!hbt]
\centering
\includegraphics[width=3in]{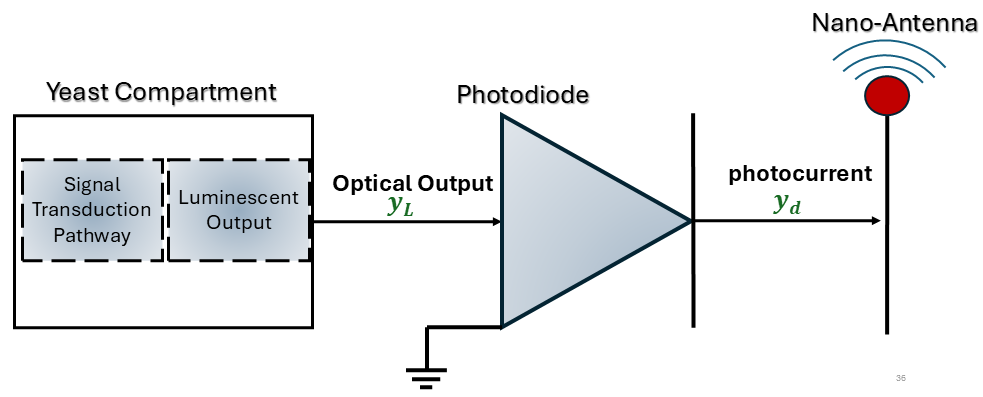}
\caption{Schematic view of the biosensor.}
\label{system description}
\end{figure}

The theoretical analysis conducted in this paper is extended to include replacement of GFP with LuxAB, where expression is driven by the \textit{CUP1} promoter and is mediated by the transcription factor Ace1. As a result, the copper detoxification pathway is coupled to the bioluminescent reporter, thereby yielding a more complete receiver model for the mathematical characterization of a yeast-derived biosensor. Key components of the proposed biosensing platform include yeast cells suspended in growth medium, coupled with a photodetector and an antenna for wireless transmission of detected events, as illustrated in Fig.~\ref{system description}.

\section{System Model} \label{system model}
The system is comprised of three main components: the yeast receiver, the photodetector, and an antenna that enables information exchange with the external environment. Interaction between the receiver and copper ions is governed by a sequence of intracellular processes, which may be classified into (i) uptake, (ii) intracellular transport, (iii) metalation and (iv) transcriptional induction. Extension of the mathematical framework to include a luminescent reporter enables an optical output to be generated. This optical signal is converted into photocurrent by a photodiode, thereby enabling subsequent wireless transmission through an antenna unit.

The primary objective is the formulation of a tractable mathematical representation of the dominant biological processes driving the receiver response, based on mechanistic information available in the literature. For the photodiode stage, established device models are adopted to characterize the output current. In vitro experiments conducted on the developed testbed consider direct Cu stimulation of yeast cells and quantify the resulting fluorescent response. As agreement between experimental measurements and theoretical predictions is achieved to a satisfactory extent, the computational model is positioned as a simulation platform through which this biosensing system may be studied from an MC perspective. In addition, virtual experiments may be enabled in cases where laboratory cost or time constraints are limiting, while mechanistic insight into the underlying biological processes may be provided.

Nevertheless, since such a nuanced model of this pathway has not been documented previously to the best of the authors’ knowledge, the presented analysis is intended to serve as a basis for further refinement. The proposed framework takes the form of coupled differential equations that characterize the time evolution of the involved chemical compounds, including the system output.

\subsection{Receiver} \label{Receiver model}
Model construction is initiated by identifying the dominant interactions among the chemical compounds participating in the copper detoxification pathway, as described in \cite{Shi2020}. Copper is assumed to be available in the cuprous form (Cu(I)) following reduction by the plasma-membrane reductases Fre1 or Fre2. Uptake is facilitated by two copper transporters (Ctr) located in the plasma membrane, namely Ctr1 and Ctr3. These form homo-trimeric, high affinity Cu-selective pores across the membrane. After internalization, Cu(I) is bound by glutathione (GSH) in the cytosol. Acting as a native chelator, GSH is then involved in copper distribution toward metallochaperones, including Atx1, Ccs1, and Cox17, through which copper is delivered to specific intracellular targets.

Ccs1 is responsible for escorting copper to Sod1 in the cytoplasm, which is associated with detoxification and redox regulation. Through Atx1, copper is delivered to Ccc2, by which copper is loaded onto additional cuproproteins (CuPrs) within the Golgi apparatus. Additional transport routes mediated by GSH include copper delivery to mitochondrial metallochaperones (e.g., Cox17, Cox11, and Sco1/2) for subsequent loading to cytochrome \textit{c} oxidase (CcO), copper sequestration into vacuoles under elevated conditions, and handoff to scavenger proteins that contribute to buffering of free copper ions.

To preserve intracellular homeostasis and prevent toxicity, transcriptional regulators are activated and mediate expression of genes involved in copper uptake and detoxification. In the system considered here, the output is provided by the expression of the \textit{CUP1} gene, which encodes the metallothionein protein Cup1. Its induction is mediated by the transcriptional regulator Ace1, following copper-dependent activation in the nucleus. Copper uptake-related genes, including \textit{CTR1} as well as the reductases \textit{FRE1}/\textit{FRE2}, are induced by the transcriptional regulator Mac1 in an analogous regulatory role. A detailed description of the copper detoxification pathway is provided in \cite{Shi2020}.

The exact mechanism through which Cu ions activate Ace1 and Mac1 via GSH remains incompletely understood. However, in \cite{Shi2020} it is noted that Cu(I)-GSH complexes have been detected in the nucleus, which suggests that a fraction of intracellular copper may be transferred directly from GSH to Ace1. Owing to the limited mechanistic detail currently available, direct loading of copper from GSH to Ace1 is assumed in this work. Cu transfer to Ace1 leads to Ace1 metallation and subsequent \textit{CUP1} induction.

\begin{figure}[!hbt]
\centering
\includegraphics[width=\linewidth]{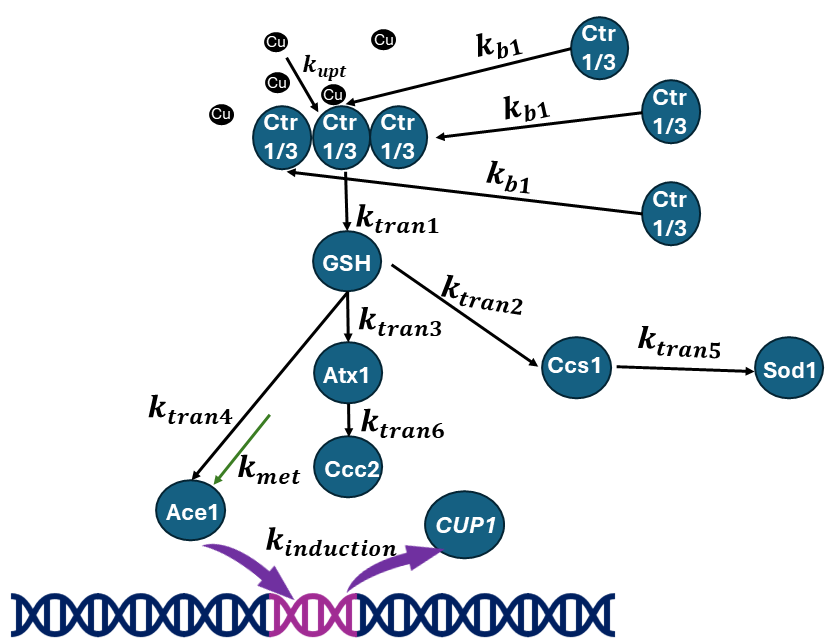}
\caption{Schematic of reaction processes within the copper detoxification pathway.}
\label{pathway schematic}
\end{figure}

Transport of copper to the vacuole and mitochondria is also not fully characterized and is therefore excluded from the present analysis. Moreover, transcriptional regulation mediated by Mac1 is omitted, while Ctr1/3 production is approximated using zero-order kinetics. As illustrated in Fig.~\ref{pathway schematic}, the reaction set considered in this work includes Ctr1/Ctr3, GSH, Atx1, Ccs1, Sod1, Ccc2, Ace1, and \textit{CUP1}.

Chemical reactions within the copper detoxification pathway considered here are grouped into five main reaction types: uptake, trimerization, transport (handoff), metalation, and transcriptional regulation. The chemical equations summarized in Table~\ref{chem eqs}, together with the reaction diagram in Fig.~\ref{pathway schematic}, provide the foundation upon which the reported mathematical models are derived. In the subsequent analysis, concentration species are denoted in brackets for readability.

\begin{table}[!t]

  \caption{Chemical Equations governing the various chemical reactions within the copper detoxification pathway.}
  \label{chem eqs}
  \centering
  \setlength{\tabcolsep}{1.5pt}
  \begin{tabular}{|p{2cm}|c|}
   \hline
Chemical reaction&Chemical Equation\\
    \hline
$Ctr1/3$ trimerization&$Ctr1/3 + Ctr1/3 + Ctr1/3 \xleftrightharpoons[k_{bind}]{k_{unbind}} Ctr1/3_{trim}$\\
    \hline
$Ctr1/3$ trimer loaded with Cu&$Ctr1/3_{trim} + Cu \xrightarrow{k_{upt}} Ctr1/3_{trim}Cu$\\
    \hline
Cu handoff from Ctr1/3 to GSH&$\begin{array}{l}Ctr1/3_{trim}Cu + GSH \\\xrightarrow{k_{tran}} Ctr1/3_{trim}Cu-GSH \\ \xrightarrow{k_{tran}} Ctr1/3_{trim} + GSHCu\end{array}$\\
    \hline
Cu handoff from GSH to Ccs1&$\begin{array}{l}GSHCu + Ccs1 \xrightarrow{k_{tran}} GSHCu-Ccs1 \\\xrightarrow{k_{tran}} GSH + Ccs1Cu\end{array}$\\
    \hline
Cu handoff from Ccs1 to Sod1&$\begin{array}{l}Ccs1Cu + Sod1 \xrightarrow{k_{tran}} Ccs1-Sod1Cu \\\xrightarrow{k_{tran}} Ccs1 + Sod1Cu\end{array}$\\
    \hline
Cu handoff from GSH to Atx1&$\begin{array}{l}GSHCu + Atx1 \xrightarrow{k_{tran}} GSHCu-Atx1 \\\xrightarrow{k_{tran}} GSH + Atx1Cu\end{array}$\\
    \hline
Cu handoff from Atx1 to Ccc2&$\begin{array}{l}Atx1Cu + Ccc2 \xrightarrow{k_{tran}} Atx1Cu-Ccc2 \\\xrightarrow{k_{tran}} Atx1 + Ccc2Cu\end{array}$\\
    \hline
Cu handoff from GSH to Ace1&$\begin{array}{l}GSHCu + Ace1 \xrightarrow{k_{tran}} GSHCu-Ace1 \\\xrightarrow{k_{metal}} Ace1_{act}\end{array}$\\
    \hline
\textit{CUP1} transcription&$\begin{array}{l}Ace1Cu + D \xrightarrow{transcr} \textit{CUP1}\end{array}$\\
    \hline
\end{tabular}
\end{table}

Copper uptake is characterized by saturable kinetics \cite{Lin1990}, approaching a plateau at elevated extracellular copper concentrations. Accordingly, the uptake rate is modeled using a Michaelis--Menten-type expression: 
\begin{equation}
v_{uptake} \propto \frac{V_{max}\,[S]}{K_D + [S]}
\label{uptake}
\end{equation}

The trimerization of the copper transporters is modeled as a third order reaction of the form: 

\begin{equation}
Ctr1/3 + Ctr1/3 + Ctr1/3 \xrightleftharpoons[k_{d1}]{k_{b1}} Ctr1/3_{trimer}\label{trimerization}
\end{equation}

The binding rate constant of the Ctr molecules is denoted by $k_{b1}$. A reverse reaction is also considered, through which trimer dissociation occurs with a rate constant $k_{d1}$, thereby reforming the initial reactants. Copper binding has been reported to induce conformational changes that disrupt Ctr-mediated transport \cite{Wu2009}. In particular, copper-dependent destabilization has been observed for Ctr1 \cite{Pena2000}. Since the corresponding dose-response kinetics have not been fully characterized, this additional loss in transportation activity is represented empirically by the Hill-type degradation rate $k_{Cu-deg}$ defined in Appendix \ref{ODEs}. Because Ctr1 and Ctr3 are represented as a common species, applying this Ctr1-derived behavior to both transporters constitutes a simplifying modeling assumption.

From a Systems Biology perspective, Cu-handoff reactions are treated as distinct from simple binding processes, since different reaction mechanics are implied by their chemical formulation. In particular, the Cu transport (handoff) process may be represented as a short chain of reactions in which the Cu carrier (denoted $A$Cu) initially binds to the Cu recipient (denoted $B$), followed by dissociation of the transient complex and formation of $A$ and $B$Cu. Such transport is suggested to be facilitated by protein--protein interactions \cite{Chacn2018,Hatori2016}. An example is provided in Eq.~\eqref{handoff}:

\begin{equation}
ACu + B \xrightarrow{k_{tranA}} ACu-B \xrightarrow{k_{tranB}} A+BCu \label{handoff}
\end{equation}

In this work, the Cu-handoff reactions are treated as irreversible, such that copper is assumed to be transferred unidirectionally from one chaperone to the next. This approximation is deemed reasonable, since in \cite{Hatori2016} copper binding rates to the involved chaperones were reported to be substantially higher than the corresponding unbinding rates, thereby suggesting predominantly forward flux. Moreover, in \cite{Chacn2018} it was observed that formation of the intermediate complex during handoff occurs rapidly, while formation of the Cu--recipient complex constitutes the rate-limiting step. Given the limited mechanistic characterization available for yeast, the same assumption is adopted here, and the intermediate complex is neglected to maintain reasonable model complexity without sacrificing biological relevance.

Mechanistic information regarding metalation of the transcription factor (TF) Ace1 remains limited. Nevertheless, in \cite{furst1989} Ace1 binding to \textit{CUP1} promoters was reported to exhibit switch-like behavior and saturation at elevated copper concentrations, resulting in a sigmoidal response. Accordingly, the metallation rate is modeled using a Hill-type expression of the form:

\begin{equation}
f_{met} \propto \frac{[Cu_{donor}]^n}{ K_{D,Ace1}^n +[Cu_{donor}]^n}
\label{metalization rate}
\end{equation}
Where $K_{D,Ace1}$ is the half-maximum constant for the metallation of Ace1 by Cu ions.

Transcriptional regulation is initiated through binding of the TF Ace1 to the \textit{CUP1} promoter, thereby enabling gene induction. Derivation of the transcription rate is based on a quasi-steady-state (QSS) assumption for binding and unbinding of Ace1 molecules to the promoter. Since Ace1 binds to the promoter in monomeric form \cite{pena1998}, a Hill-type expression with $n=1$ would be obtained for the fractional transcription level which implies the probability that the promoter is occupied by the TF. However, Ace1 is known to bind to four Metal Response Elements (MREs) within the \textit{CUP1} promoter. Therefore, following an approach similar to \cite{Mitre2016}, the activation function based on the transcription probability is modeled as:

\begin{equation}
f_{act} \propto \frac{1}{\sum_{n=1}^{4} (K_{D,mRNA}/[Ace1])^n+1}
\label{fractional transcription}
\end{equation}

Where $K_{D,mRNA}$ is the TF-promoter dissociation factor. The derivation of Eq.~(\ref{fractional transcription}) is given in Appendix \ref{fact derivation} in more detail.

\subsubsection{Lux Model}\label{Bioluminescence}

The LuxAB enzyme catalyzes the light-emitting reaction by utilizing reduced flavin mononucleotide ($\mathrm{FMNH_2}$), molecular oxygen ($\mathrm{O_2}$), and a long-chain aldehyde ($\mathrm{RCHO}$) to produce oxidized FMN, a fatty acid ($\mathrm{RCOOH}$), water, and light, as shown below:

\begin{align} \label{light pathway}
FMNH_2+RCHO+O_2 \xrightarrow{\redcircle{LuxAB}} FMN+&RCOOH+ \nonumber\\H_2O+ \bluecircle{Light}
\end{align}

The substrates required for LuxAB catalysis are shown on the left-hand side of Eq.~\eqref{light pathway}. Oxidized Flavin Mononucleotide ($\mathrm{FMN}$) participates in the reaction catalyzed by the Flavin Reductase P (FRP) in the presence of the cofactors Nicotinamide adenine dinucleotide (NADH) or Nicotinamide adenine dinucleotide phosphate (NADPH), through which $\mathrm{FMNH_2}$ is produced. In parallel, the LuxCDE multi-enzyme reductase complex converts fatty acyl-ACP into $\mathrm{RCHO}$ required by LuxAB. In this process, LuxD acts as a transferase that generates fatty acids from either fatty acyl-ACP or acyl-CoA. The resulting fatty acid is ATP-activated to form acyl-AMP, which is subsequently attached to LuxE to form an acyl intermediate. It is believed that this intermediate is channeled to LuxC, where an additional acyl intermediate is formed and is reduced by NADPH to yield the aldehyde.

The light-emission pathway proceeds through multiple intermediate stages, beginning with binding of $\mathrm{FMNH_2}$ and oxygen to LuxAB and formation of the Lux:FMNHOOH complex (C4a-hydroperoxyflavin). As discussed in \cite{TINIKUL2020}, deprotonation of this C4a adduct leads to formation of C4a-peroxyflavin, which can attack the aldehyde and yield the Lux:FMNHOH:RCOOH complex, also referred to as excited C4a-hydroxyflavin. This intermediate is commonly treated as the light-emitting species and relaxes to the ground state through photon emission at approximately 490~nm. The resulting non-excited hydroxyflavin undergoes dehydration, after which fatty acid and FMN are released.

\begin{figure}[!hbt]
\centering
\includegraphics[width=\columnwidth]{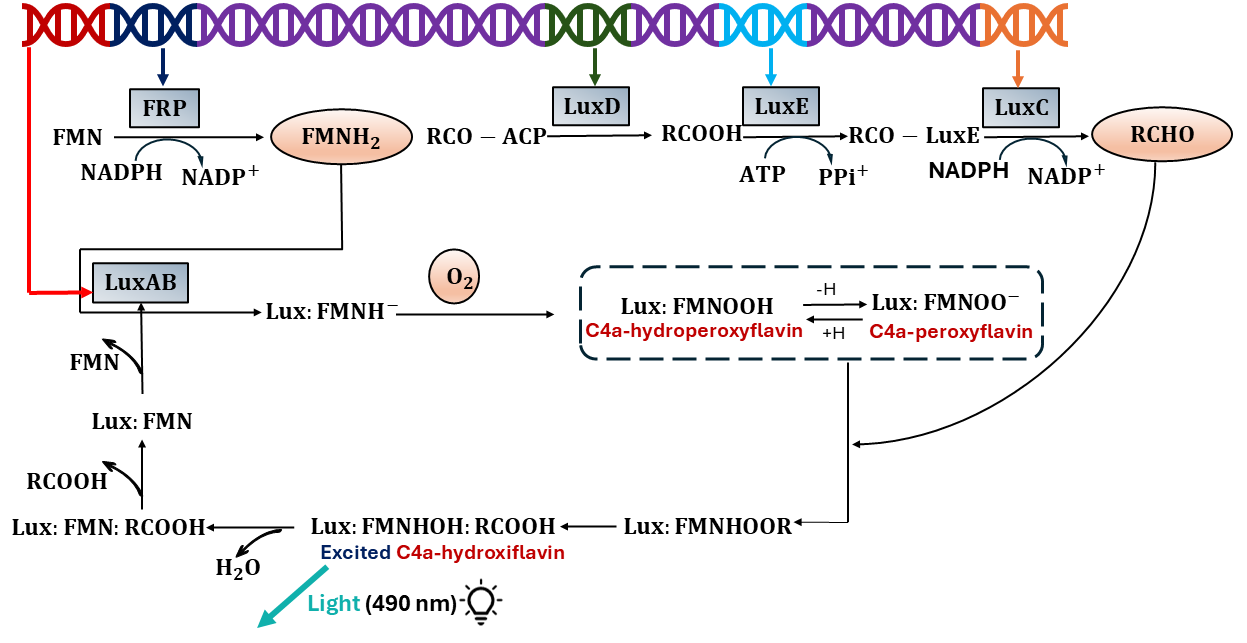}
\caption{Light pathway.}
\label{light cycle}
\end{figure}

Recycling of the two key substrates required for LuxAB catalysis enables self-sustained bioluminescent reporting without external substrate supplementation. This is suitable for MC-integrated biosensing applications targeting health monitoring through Smart Pill technologies. The light pathway incorporated into the mathematical model is illustrated schematically in Fig.~\ref{light cycle}.

The enzymatic reactions described above and depicted in Fig.~\ref{light cycle} involve more than one substrate, and therefore depart from classical kinetic formulations such as Michaelis--Menten models. For the derivation of rate equations that are suitable for ODE models describing each enzymatic reaction, the King--Altman schematic method is adopted \cite{cornish2013}. The King--Altman approach is applicable to mechanisms in which an arbitrary number of substrates bind to an enzyme in various arrangements, while a steady-state assumption is applied to the enzyme forms that are created. Multi-enzyme reactions can also be handled using this method.

Starting from basic principles, a single enzyme--multiple substrate reaction is considered with $n$ substrates available. The $n$ different enzyme forms that are created are denoted as $E_1, E_2,...E_n$. The rate of change of a species $\mathrm{E_i}$ can be given by $\frac{d[E_i]}{dt}=k_{1i}[E_1] + k_{2i}[E_2]+...+k_{ni}-\sum k_{ij}[E_i]$. Here, the rate constant for $E_i\rightarrow E_j$ is denoted by $k_{ij}$, and the rate constant for $E_j\rightarrow E_i$ is denoted by $k_{ji}$. Moreover, total enzyme concentration is conserved:
\begin{equation}
[E_1]+[E_2]+...+[E_n]=[E_{tot}]
\label{total enzyme}
\end{equation}
We solve for the m-th enzyme form concentration in a system of equations where the m-th equation is replaced with Eq. (\ref{total enzyme}):
\begin{align}\label{SS system}
    -\sum k_{1j}[E_1] + k_{21}[E_2] +...+k_{m1}[E_m] +...+ k_{n1}[E_n] = 0& \nonumber\\
    k_{12}E_1 - \sum k_{2j}[E_2] +...+k_{m2}[E_m] + ... + k_{n2}[E_n] = 0&\nonumber\\
    \vdots& \nonumber\\
    [E_1] + [E_2] +...+ [E_m] +...+ [E_n] =[E_{tot}]&\nonumber\\
    \vdots \nonumber\\
    k_{1n}[E_1] + k_{2n}[E_n] +...+ k_{mn}[E_m] +...-\sum k_{nj}[E_n] = 0&\nonumber\\
\end{align}
To solve for [$E_m$], Cramer's rule is applied, resulting in the following expression:
\begin{equation}\label{Em}
[E_m] = \frac{N_m\,[E_{tot}]}{D}
\end{equation}
Where the numerator $\mathrm{N_m}$ is given by the following determinant:
\begin{equation}\label{Nm}
\setlength{\arraycolsep}{2pt}
\resizebox{\linewidth}{!}{$
N_m=
\begin{vmatrix}
-\sum_j k_{1j} & k_{21}         & \cdots & k_{n1} \\
k_{12}         & -\sum_j k_{2j} & \cdots & k_{n2} \\
\vdots         & \vdots         & \ddots & \vdots \\
k_{1n}         & k_{2n}         & \cdots & -\sum_j k_{nj}
\end{vmatrix}
$}
\end{equation}
And the denominator $D$:

\begin{equation}\label{D}
\setlength{\arraycolsep}{2pt}
\resizebox{\linewidth}{!}{$
D=
\begin{vmatrix}
k_{12} & -\sum_j k_{2j} & \cdots & k_{m2} & \cdots & k_{n2} \\
\vdots & \vdots         &  & \vdots &        & \vdots \\
1      & 1              & \cdots & 1      & \cdots & 1 \\
\vdots & \vdots         &        & \vdots &  & \vdots \\
k_{1n} & k_{2n}         & \cdots & k_{mn} & \cdots & -\sum_j k_{nj}
\end{vmatrix}
$}
\end{equation}

Expansion of the determinant in Eq. (\ref{Nm}) reveals some key properties which form the basis for the King-Altman schematic method. Every product in the expansion of $N_m$ contains (n-1) constants $k_{ij}$, essentially forming reaction "paths" among the different enzyme forms. The index m cannot occur as a first index in any $k_{ij}$ within a product. Every other index can occur once as first index in every product, while cycles of indices are canceled within the expansion. Ultimately, every product that is left in $N_m$ has the same sign, where $m$ occurs at least once as a second index (every "reaction path" that survive in $N_m$ contain $E_m$ as a "sink"). 

It is assumed in the general derivation that all possible enzyme forms react directly with each other; however, such full connectivity is typically not realized in practice. Non-feasible reactions are eliminated by setting the corresponding rate constants to zero, whereby the associated paths are removed from $N_m$. The denominator $D$ is obtained as the sum of all numerators, i.e., $D = N_1+N_2+..N_m+...N_n$. Further theoretical details on the principles summarized above are provided in \cite{cornish2013}.

Application of the principles outlined above to enzymatic reactions such as those depicted in Fig.~\ref{light cycle} yields schematic representations that are subsequently used for derivation of rate equations to be included in the receiver ODE model. Enzymatic reaction schematics are patterns that connect the different enzyme forms according to the binding and release events dictated by the underlying physical mechanism. In multi-substrate reactions, different substrate-binding schemes may arise, primarily depending on the order of binding events \cite{cornish2013}, with common mechanisms including compulsory-order and random-order binding. Based on the available information, the enzymatic reactions comprising the light pathway are assumed here to follow compulsory-order kinetics.

\begin{figure}[!hbt]
\centering
\includegraphics[width= 2.2in]{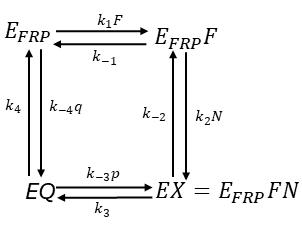}
\caption{Master pattern for sequential binding of FMN and NADPH to FRP.}
\label{FRP master pattern}
\end{figure}

The reaction between FRP, FMN, and NADPH is modeled as ternary-complex formation involving two substrate-binding steps, namely $E_{FRP}+F\xrightleftharpoons[k_{-1}]{k_1}E_{FRP}F$ and $E_{FRP}F + N\xrightleftharpoons[k_-{2}]{k_2}E_{FRP}FN$, followed by two product-release steps, $EPQ\xrightleftharpoons[k_-{3}]{k_3}EQ+P$ and $EQ\xrightleftharpoons[k_-{4}]{k_4}E+Q$. Here, FMN and NADPH are denoted by F and N, respectively. The intermediate $E_{FRP}FN$ is transformed to $EPQ$ through $E_{FRP}FN\rightleftharpoons EPQ$, where $P$ and $Q$ denote the two products, namely $\mathrm{FMNH_2}$ and $\mathrm{NADP^+}$, which are released. Due to limited information regarding intermediate states between $E_{FRP}FN$ and $EPQ$, these two complexes are treated as a single species and are collapsed into a single complex $EX$. As part of the King--Altman method, reactions are expressed using pseudo first-order kinetics, as illustrated by the master pattern in Fig.~\ref{FRP master pattern} (e.g., the bimolecular rate constant $k_1$ is replaced by the pseudo first-order rate constant $k_1F$).
The next step of the schematic method is to derive all possible schemes based on the master pattern.
\begin{figure}[!hbt]
\centering
\includegraphics[width= 2.5in]{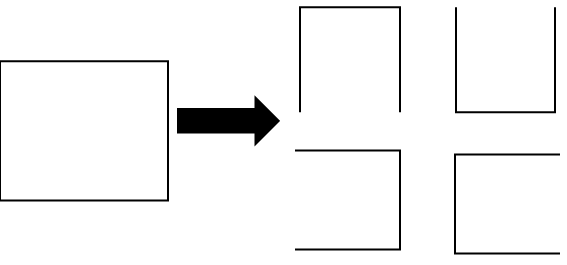}
\caption{All possible King-Altman Patterns.}
\label{patterns}
\end{figure}
The velocity of the first product release (P), namely $\mathrm{FMNH_2}$, is in this case:\newline
\begin{align}\label{EX vel1}
    \frac{d[P]}{dt}\propto \upsilon_{_{FMNH_2}}=k_3\,[EX]-k_{-3}p\,[EQ] = k_3\frac{N_{EX}[E^{FRP}_{tot}]}{D}-\nonumber\\k_{-3}p\frac{N_{EQ}[E^{FRP}_{tot}]}{D}
\end{align}

\begin{figure}[htpb]
  \centering
  \subfloat[Pattern a.]{\includegraphics[width=1.5in]{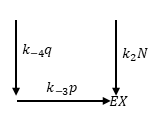}%
 \label{patt a}}
  \hfill
  \subfloat[Pattern b.]{\includegraphics[width=1.3in]{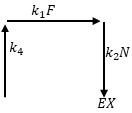}%
 \label{patt b}}
 \hfill
   \subfloat[Pattern c.]{\includegraphics[width=1.3in]{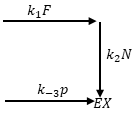}%
 \label{patt c}}
 \hfill
    \subfloat[Pattern d.]{\includegraphics[width=1.5in]{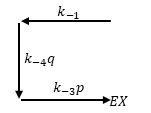}%
 \label{patt d}}
 \hfill
  \caption{Legitimate King-Altman patterns for the flavin reduction reaction.}
  \label{legitimate patterns}
\end{figure}

Product inhibition is not considered in this work; therefore, product release is treated as non-reversible (i.e., $k_{-3}$ and $k_{-4}$ are set to zero), and the corresponding rate derivations pertain to initial-rate conditions. In this case, it is easy to prove that $\frac{dP}{dt}=\frac{dQ}{dt}$ (equal product release rate), and thus $\upsilon_{_{FMNH_2}}=\upsilon_{_{NADPH}}$, using the steady state property for enzyme forms in a King-Altman setting. Based on the patterns depicted in Fig.~\ref{patterns}, together with the principles outlined above, the legitimate King--Altman patterns for this reaction are obtained as shown in Fig.~\ref{legitimate patterns}.

The only pattern that does not contain product inhibition, is the one depicted in Fig. \ref{patt b}. Therefore, it contributes to $N_{EX}$ which is given by the following expression:
\begin{equation}\label{N_EX}
    N_{EX} = k_4\,k_1[F]\,k_2[N]
\end{equation}

By deriving the legitimate King--Altman patterns for the remaining enzyme forms involved in this reaction, namely $E_{FRP}$, $E_{FRP}F$, $E_{FRP}FN$, and $EQ$, the denominator $D_{FRP}$ can be constructed and the full expression in Eq.~(\ref{EX vel1}) can be obtained. Subsequently, Cleland’s notation is adopted to group rate constants, as described in \cite{cornish2013}, \cite{segal1976}, thereby transforming the result into a more convenient form:

\begin{equation}\label{Cleland FMN}
    \upsilon_{_{FMNH_2}}=\frac{V^+_{FRP}\,[F]\,[N]}{D_{FRP}}
\end{equation}
And,
\begin{equation}\label{Dfrp}
D_{FRP} = K_{D,F}\,K_{m,N}+K_{m,N}[F]+K_{m,F}\,[N]+[F]\,[N]
\end{equation}
Where the forward catalytic rate is defined as $V^+_{FRP}=k_{cat}\,[FRP]$, the Michaelis constants are denoted by $K_{m,F}$ and $K_{m,N}$ for FMN and NADPH respectively, and $K_{D,F}$ represents the dissociation constant of FMN. The derivation of Eq.~(\ref{Cleland FMN}) is provided in Appendix~\ref{KA Method}, together with the rest of the enzymatic rates presented here. The conversion of fatty acyl-ACP to aldehyde by the reductase LuxCDE is treated in a similar fashion. It is further assumed that fatty acid is recycled entirely as one of the products of the reaction between LuxAB, $\mathrm{FMNH_2}$, RCHO, and $\mathrm{O_2}$; therefore, the aldehyde supply chain is assumed to be realized solely through LuxE and LuxC. Hence, the acyl-transferase LuxD is omitted by the mathematical model. The velocity of the reaction between RCOOH, ATP and LuxE for the production of acyl-LuxE (LuxE-RCO), can be given in a similar fashion as follows:

\begin{equation}\label{Cleland RCOOH}
    \upsilon_{_{LuxE-RCO}}=\frac{V^+_{LuxE}\,[RCOOH]\,[ATP]}{D_{LuxE}}
\end{equation}
Where 
\begin{align}\label{D_RCOOH}
    D_{LuxE} = K_{D,RCOOH}\,K_{m,ATP} +K_{m,ATP}\,[RCOOH]&+\nonumber\\K_{m,RCOOH}\,[ATP]+[RCOOH]\,[ATP]
\end{align}
The velocity of the reaction between LuxC, LuxE-RCO and NADPH for the production of aldehyde (A), is given by:

\begin{equation}\label{Cleland RCHO}
    \upsilon_{_{A}}=\frac{V^+_{LuxC}\,[LuxE-RCO]\,[N]}{D_{LuxC}}
\end{equation}
Where 
\begin{align}\label{D_LuxC}
D_{LuxC}=K_{D,LuxE-RCO}\,K_{m,N2}+\nonumber\\K_{m,N2}\,[LuxE-RCO]+&\nonumber\\K_{m,LuxE-RCO}\,[N]+[LuxE-RCO]\,[N]
\end{align}

Catalysis by LuxAB is described by the processes shown in Fig.~\ref{light cycle}. Beyond substrate-binding and product-release events, multiple intermediate stages are involved, which play a critical role in both light emission and substrate recycling \cite{TINIKUL2020}. Owing to the distinct nature of this light pathway, the corresponding King--Altman schematic becomes more complex and requires special treatment.

\begin{figure}[!hbt]
\centering
\includegraphics[width=\linewidth]{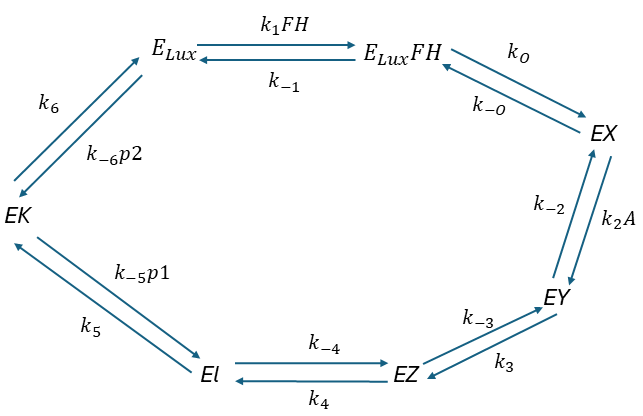}
\caption{King-Altman pattern for the catalysis of LuxAB and light emission.}
\label{lux KA}
\end{figure}

As shown in Fig.~\ref{lux KA}, the first enzyme--substrate complex is denoted by $E_{Lux}\,FH$, where FH represents the substrate $\mathrm{FMNH_2}$. Upon oxygen binding, a ternary complex is formed and the intermediate $EX$ is generated. Here, $\mathrm{O_2}$ is assumed to be constant and is therefore absorbed into the pseudo first-order constant rate $K_O$. The intermediate $EX$ represents a critical state of the system, since it corresponds to the enzyme form that undergoes deprotonation prior to binding of the aldehyde substrate (denoted by $A$). Since protonation and deprotonation typically occur on fast timescales, thermodynamic equilibrium between the protonated and deprotonated state of $EX$ (see Fig.~\ref{light cycle}) is commonly assumed \cite{cornish2013}. As such, the aldehyde-binding rate constant $k_{bA}$ is replaced by:

\begin{equation}\label{Aldehyde bind}
    k_2=k_{bA}\,\frac{K_E}{K_E+[h]}
\end{equation}
where the equilibrium constant is denoted by $K_E$ and the cytosolic hydrogen concentration by [h]. The complex denoted by $EY$ corresponds to the aldehyde-bound intermediate (flavin C4a-peroxyhemiacetal), which is subsequently transformed into $EZ$. The complex $EZ$ represents the excited C4a-hydroxyflavin state and is treated as the light-emitting species. The intermediate $EL$ corresponds to the dehydrated/de-excited complex following photon emission. Here, $\mathrm{H_2O}$ release is assumed to be constant and its effect is absorbed into $k_4$, leading to $EK$, where fatty-acid release has occurred. Subsequent release of $\mathrm{FMN}$ returns the system to the free enzyme form $E_{Lux}$. Photon emission is given by:

\begin{equation}\label{light eq}
    [Ph]=k_L\,[EZ]
\end{equation}
Where $k_L$ is the photon emission rate constant in $min^{-1}$, [$Ph$] is computed in $\mu mol$ photons $L^{-1}\,min^{-1}$, and $[EZ]$ is computed in   $\mu mol\,L^{-1}$. The derivation of $[EZ]$ is provided in Appendix \ref{KA Method}. The number of photons per minute can be given by the following formula:

\begin{equation}\label{number of photons}
    n_L=10^{-6}N_A\,[Ph]\,V_{cell}
\end{equation}
where $N_A$ is the Avogadro number and $V_{cell}=1.13\cdot10^{-13}$ liters denotes the yeast cell volume considered here \cite{Milo2009}. The full ODE model, together with the parameter values and initial concentrations used in simulations, is provided in Appendix~\ref{ODEs}, where all concentration values are expressed in $\mu$M and time is depicted in minutes.

Due to the novelty of the proposed model and the scarcity of yeast-specific quantitative data, parameter values were selected from the yeast literature whenever possible. When relevant yeast data were unavailable, parameter values reported for bacterial processes were adopted. When no information was available, reasonable estimates were assumed based on acceptable value ranges for the corresponding processes. The initial concentration values that are not mentioned in Appendix~\ref{ODEs}, are set to zero.

\subsection{Photodetector}\label{Photodetector}
The photocurrent output of a photodetector coupled to the receiver model described in Section~\ref{Receiver model} is derived using the photon emission rate from the Lux reporter, as defined in Section~\ref{Bioluminescence}. In this work, a CMOS p+/n-well/psub  photodiode akin to the one used in \cite{Koklu2013} is considered as the photodetector element, consistent with sensing circuitry employed in prominent biosensors such as Smart Pills \cite{IndaWebb2023}. The corresponding photocurrent response can be expressed as follows:
\begin{equation} \label{photocurrent}
    I_{pd} = R_{pd}\,P_L
\end{equation}
where $R_{pd}$ denotes the photodiode responsivity in Amperes per Watt ($A/W$) and $P_L$ denotes the optical power generated by yeast bioluminescence in $W$. The optical power is obtained by combining the photon number (per second) with the Planck constant $h_p$, the speed of light $C$ and the emission wavelength $\lambda$ (in meters) as follows:

\begin{equation} \label{optical power}
P_L = \frac{n_L}{60}\,\frac{h_p\,C}{\lambda}
\end{equation}
For the Lux-mediated light emission, $\lambda=490$ nm.

\section{Experimental Setup}\label{testbed}
Reliability of the mathematical tools presented in Section~\ref{system model} depends on the availability of an experimental platform through which theoretical findings can be validated. In the context of advancing MC toward biosensing applications, such a setup provides a foundation for development of yeast-based MC receivers that are suitable for integration within biosensors. Moreover, the regulatory networks governing metal homeostasis in \textit{S. cerevisiae} can be leveraged as a basis for a broad class of engineered yeast biosensors. Accordingly, the testbed presented in this work is informed by the extensively studied yeast response to copper stress, with particular emphasis placed on sequestration and detoxification mechanisms under heavy-metal accumulation \cite{Tohoyama1995}.

\subsection{Testbed Description}\label{testbed description}
As described in Section \ref{description}, WT yeast cells feature an intrinsic biological property that tightly controls intracellular copper concentrations, thereby maintaining metal homeostasis under physiological conditions. Copper, despite being an essential micronutrient for the yeast physiology can become toxic at excessive concentrations \cite{Kim2008}, and triggers inherent detoxification mechanisms which restore cellular homeostasis \cite{Butt1984}. Translocation of the copper-responsive transcriptional regulator Ace1 controls the induction of the \textit{CUP1} gene which encodes the metal-binding Cup1 metallothionein to discharge the cell from cytotoxic buildup \cite{Butt1984}, \cite{thiele1988}.

The experimental platform reported here leverages the well-established pathway of copper-inducible \textit{CUP1} gene activation, to engineer a fluorescent yeast "receiver" capable of sensing the presence of accumulated copper. This has been implemented by replacing the \textit{CUP1} gene with the \textit{GFP} gene, enabling the expression of a green fluorescent protein under the control of the copper-responsive promoter \textit{pCUP1} (Section~\ref{description}, Fig.~\ref{gfp yeast}). To this end, the engineered receiver is predicted to emit a green signal when exposed to high copper concentrations, providing a measurable output for MC. The overall system is foreseen to deepen our understanding on the dissemination of information through molecular signaling.
\subsection{Materials and Methods}
\subsubsection{Yeast strains and cultivation conditions}
All yeast strains in this study were originated from \textit{S. cerevisiae} BY4741 ('wild type', genotype MATa) and transformed with PCR products by the lithium acetate method \cite{Gietz2002}. The $cup1\Delta$::sfGFP reporter strain was constructed by genetic replacement of the \textit{CUP1-2} gene via homologous recombination of a PCR product containing the superfolder GFP (sfGFP) fused to a CLN2 PEST sequence and the natMX cassette for selection. The \textit{CUP1-1} gene locus remained intact in the yeast genome. The constructed strain was verified through DNA sequencing. The PCR product was amplified from the Gal1pORF-sfGFPdeg-natMX strain \cite{Bheda2020}. All oligonucleotides used for PCR-amplifications are listed in Table \ref{oligonucleotides}.
Yeast cells were cultivated to saturation overnight at 30 $^\circ$C in selective YPAD medium (0.12 g/L adenine hemisulfate salt, 10g/L yeast extract, 20 g/L peptone and 20 g/L glucose) supplemented with appropriate antibiotics. The next day, cells were diluted to $\mathrm{OD_{600}=0.2}$ and grown for a few hours until $\mathrm{OD_{600}}$ reached approximately 0.6, prior to experimental induction.
\subsubsection{Copper induction and cell harvesting}
Yeast cells at the logarithmically growing phase ($\mathrm{OD_{600}\sim0.6}$) were stimulated with either 1, 10, 100 or 500 $\mathrm{\mu M}$ of copper sulfate (Sigma-Aldrich) and samples of cell suspension were harvested at specific time points following initial stimulation. 
For the RNA assay, aliquots of $\sim 3\cdot 10^6$ cells were harvested per each time point, washed with sterile water and stored at -80 $^\circ$C before proceeding with RNA isolation.
For fluorometric measurements, stimulated cells were adjusted to the exponential phase of growth, washed once with sterile water and aliquots of $\sim 3\cdot 10^6$ cells were transferred to a 96-well microplate (Thermo Scientific) to monitor fluorescence yield. For the specificity assay, yeast cells were stimulated with 100 $\mu$M zinc chloride (Sigma-Aldrich). Green emission was determined in arbitrary units (AU) using the Infinite F200 fluorescence microplate reader (Tecan Trading Ltd, Switzerland) with excitation, 485/20 nm; emission, 510/20 nm; manual gain, 100. The presented data from the fluorescence assays were obtained from at least three independent experiments.
\subsubsection{RNA isolation, reverse transcription and quantitative PCR}
Total RNA was isolated from frozen samples using the hot phenol method \cite{Schmitt1990} and cDNA synthesis was performed with the PrimeScript RT Reagent Kit (Perfect Real Time, TaKaRa Bio). Quantitative real-time PCR was performed on a CFX96 thermal cycler (BioRad) using the KAPA SYBR FAST qPCR Master Mix (KAPA Biosystems). Actb ($\mathrm{\beta}$)-actin was used as a housekeeping gene and all oligonucleotide primers used in the assay are listed in Table \ref{oligonucleotides}. Relative changes in gene expression were analyzed using the $\mathrm{2^{-ΔΔC[T]}}$ algorithm. All samples and standards were run in triplicates and the presented results were obtained from at least two independent experiments.

\begin{table}[!t]
\caption{Oligonucleotide primers used in the study.}
\label{oligonucleotides}
\centering
\setlength\tabcolsep{1.5pt}
\begin{tabular}{|c||c|}
\hline
A. For PCR amplification &\\
\hline
Yeast Strain & Primer pair\\
\hline
\textit{cup1$\Delta$}::sfGFP & 
\begin{tabular}{c} F: 5’- AAGAAAAACAAACTGTACAA\\TCAATCAATCAATCATCACAT
\\AAAatgtccaagggtgaagagct -3’\\ 
\hline 	R: 5’- TTAAAAACAGCAAATAGTTAG\\ATGAATATATTAAAGACTATTC
\\GTTcagtatagcgaccagcattc -3’
\end{tabular}\\
\hline
B. For quantitative PCR &\\
\hline
Gene & Primer pair\\
\hline
sfGFP &
\begin{tabular}{c} F: 5’- CCATTTTGGTAGAACTGGAC \\-3’\\ \hline R: 5’- CATATGGTCTGGGTATCTTG \\-3’
\end{tabular}\\
\hline
$\beta$-actin &
\begin{tabular}{c} F: 5’- CATCTTCCATGAAGGTCAAG \\-3’\\ \hline R: 5’- CTTGTGGTGAACGATAGATG \\-3’
\end{tabular}\\
\hline
\end{tabular}
\end{table}

\section{Results}\label{results}
In this section, results obtained from both experimental measurements and computational analysis are presented. Experimental trials incorporate GFP-based quantification at the transcript (mRNA) and protein levels, where copper is externally supplied to the growth medium. Joint observation of these two response modalities provides insight into the dynamics of the yeast receiver. Transcript induction is being characterized across multiple copper concentrations in order to observe any dose-dependent differences in promoter activity, while fluorometric measurements are limited to a single input concentration. Protein-level dynamics are inherently slower due to additional post-transcriptional mechanisms involved and therefore, can potentially misrepresent nuances from diverse input conditions.

Theoretical results are reported for single copper-stimulation scenarios and are presented for both the GFP and Lux reporters. GFP-based computations are included primarily for validation, whereas Lux-based computations enable coupling of the receiver model to the photodetector response, thereby supporting biosensing scenarios such as event reporting in Smart Pill applications. A direct Cu input is assumed in the current model, and molecular propagation effects are therefore neglected; incorporation of propagation phenomena is left for future work.

\subsection{GFP abundance}\label{results abundance}
\begin{figure}[!hbt]
\centering
\includegraphics[width=3in]{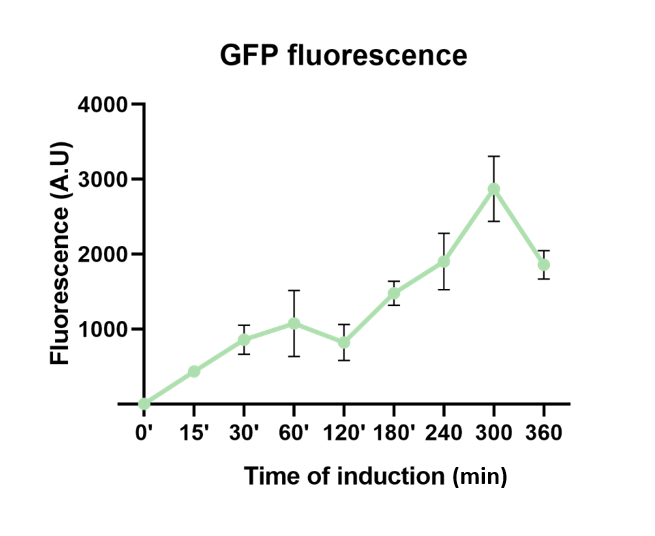}
\caption{Expression of enhanced green fluorescent protein in $cup1\Delta$::sfGFP cells following induction with 100 $\mu$M of copper sulfate. Green fluorescence intensity was monitored in copper-induced cells for 360 minutes following initial induction. Fluorescence values at each time point were normalized to those not exposed to copper sulfate. Values from three independent experiments are presented as mean $\pm$ SEM. AU, arbitrary units.}
\label{gfp protein}
\end{figure}

GFP fluorescence was monitored in $cup1\Delta$::sfGFP cells at regular intervals (as indicated on the graph) for a total duration of 6h after initial stimulation with 100 $\mathrm{\mu}$M of copper sulfate. As shown in Fig. \ref{gfp protein}, a  significant increase of green signal was detectable on a fluorescence reader as early as 15 minutes. Maximum fluorescence was reached at 300 minutes, followed by a gradual decrease in signal intensity, indicating the onset of protein turnover. The translation delay time together with the relatively slow decay of GFP due its stability, results in a slow overall response.

\subsection{RNA levels of GFP} \label{results mRNA}
Although the GFP reporter provides a measurable output of the yeast receiver in real time, the protein-level dynamics are comparatively slow due to translation and maturation delays. The demonstrated stability can provide a robust readout and reduce susceptibility to transient fluctuations. However, the overall slow nature of protein abundance reporting can limit the achievable data rate in MC scenarios where repeated stimulation of the receiver is required.

\begin{figure}[!hbt]
\centering
\includegraphics[width=3in]{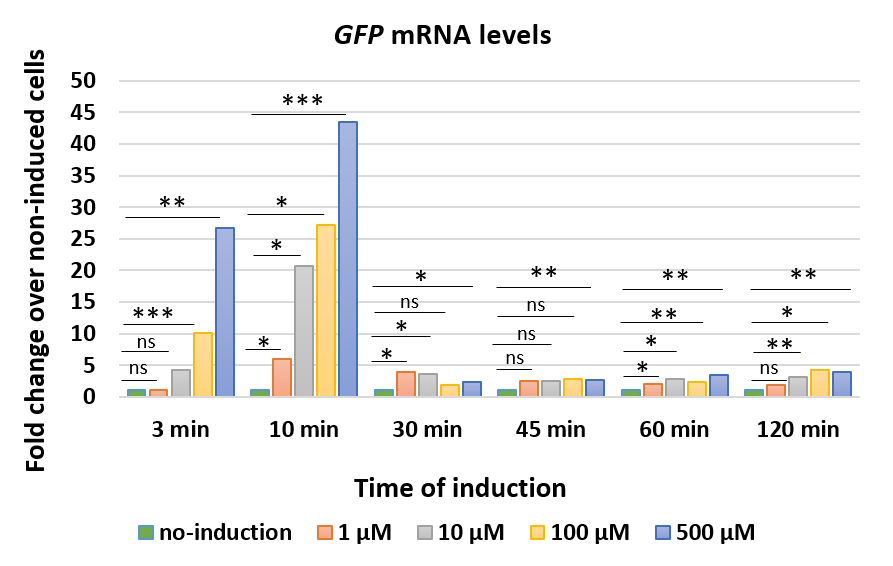}
\caption{GFP transcript levels were determined by quantitative PCR following the induction of cup1$\Delta$::sfGFP with 1, 10, 100 and 500 $\mu $M of copper sulfate as indicated in the charts. RNA levels were monitored for 120’ following initial induction at the indicated time points. Across all copper concentrations, gene expression peaked at 10 minutes, with higher copper concentrations eliciting a more robust response. GFP expression levels decreased at subsequent time points, indicating a transient induction following copper exposure. Values were normalized to the expression of $\beta$-actin and are presented as fold change over non-induced cells. Data represent mean $\pm$ SEM from two independent experiments (N=2). Statistical significance was determined using a two-tailed Student’s t-test. *$p < 0.05$, **$p < 0.01$, ***$p < 0.001$, ns – non significant.}
\label{gfp mRNA}
\end{figure}

To better elaborate on the promoter activity during induction, GFP mRNA levels were additionally measured. The single concentration used for protein abundance measurements is extended to a set of values, aiming to characterize the promoter's activity across different input conditions. Therefore, the investigation focuses on mRNA responses for copper concentrations of 1,10,100 and 500 $\mu M$. The distinct dynamics shown in Fig.~\ref{gfp mRNA} indicate that substantially faster responses can be achieved at the transcript level compared to protein abundance. 

Following stimulation with 1~$\mathrm{\mu}$M, mRNA transcripts are barely detectable-yet peak at 10 minutes, while 10~$\mathrm{\mu}$M leads to notable upregulation increases at both 3 and 10 minutes after stimulation, without altering the peak time of the response. Copper inputs of 100 and 500~$\mathrm{\mu}$M, cause a notably faster activation of the CUP1 promoter leading to detectable transcripts within 3 minutes. Expression levels peak at 10 minutes followed by a sharp decrease. Overall, the detoxification response maintains a similar transient pattern across all tested concentrations, being particularly fast-acting at higher copper dosages. The natural dynamics of the activity at the promoter level are indicative of the cellular protection against elevated copper concentrations.
Real-time RNA sequence detection has been reported in the literature \cite{Guet2015}, \cite{Kukhtevich2022}. These studies indicate that genetically encoded RNA aptamers capable of specifically binding an exogenously supplied fluorophore have emerged as a promising strategy for the precise tracking and quantification of transcripts in living cells. Among them, the Spinach aptamer has been successfully applied to label RNA molecules in living \textit{S. cerevisiae} cells. More specifically, the Spinach sequence can be cloned between the coding region and the 3′-UTR of a target gene, producing green fluorescence upon interaction with DFHBI, a GFP chromophore analogue. In parallel, advanced imaging technologies have been developed to provide real-time spatiotemporal and quantitative characterization of both highly and weakly expressed transcripts. This is of great importance, as it points to the potential of leveraging the rapid RNA response of yeast cells to achieve high event rates. 



\subsection{Biosensor Selectivity} \label{selectivity}

\begin{figure}[!hbt]
\centering
\includegraphics[width=3.5in]{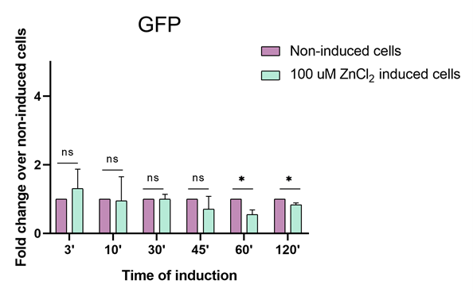}
\caption{GFP transcript levels in cup1$\Delta$::sfGFP cells following exposure to 100 $\mu$M $\mathrm{ZnCl_2}$ for the indicated induction times, determined by quantitative PCR. Zinc treatment did not induce GFP transcript expression above basal levels. Values were normalized to the expression of $\beta$-actin and are presented as fold change over non-induced cells. Data represent mean $\pm$ SEM from two independent experiments (N=2). Statistical significance was determined using a two-tailed Student’s t-test. *p < 0.05, ns – non significant.}
\label{specificity}
\end{figure}

\begin{figure}[!hbt]
\centering
\includegraphics[width=3in]{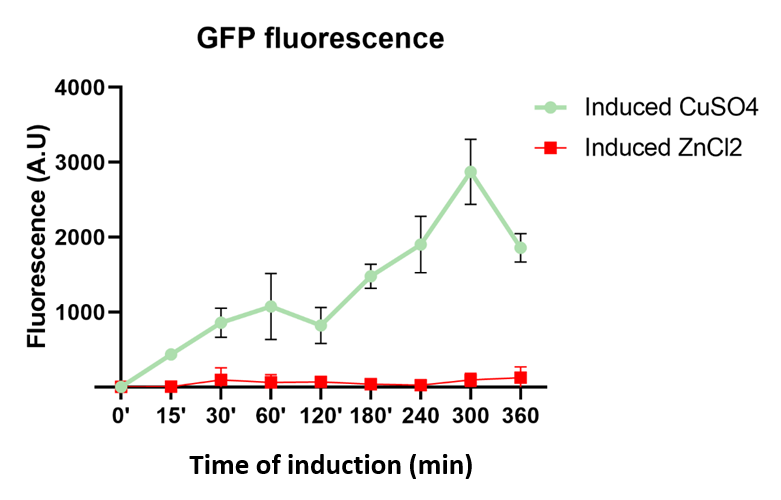}
\caption{GFP fluorescence in cup1$\Delta$::sfGFP cells following induction with 100 $\mu$M $\mathrm{CuSO_4}$ or 100 $\mu$M $\mathrm{ZnCl_2}$. While $\mathrm{CuSO_4}$ induced a marked increase in GFP fluorescence, $\mathrm{ZnCl_2}$ failed to elicit a detectable response throughout the induction period, demonstrating the specificity of the CUP1/Ace1 regulatory system for copper under the conditions tested. Data represent mean $\pm$ SEM from three independent experiments (N=3). }
\label{specificity2}
\end{figure}

To evaluate potential cross-reactivity with other metal ions, cells were exposed to $\mathrm{ZnCl_2}$, a chemically related metal that could potentially interfere with biosensor activity, under the same induction conditions used for copper stimulation. RT-qPCR analysis revealed no significant increase in GFP transcript levels following $\mathrm{ZnCl_2}$ treatment compared with non-induced controls, demonstrating that the biosensor is selectively activated by copper and not by zinc (Fig.~\ref{specificity}). Consistent with these findings, monitoring of GFP fluorescence over time showed that $\mathrm{ZnCl_2}$ failed to induce reporter expression, whereas $\mathrm{CuSO_4}$ elicited a robust and time-dependent increase in GFP fluorescence (Fig.~\ref{specificity2}). Together, these results demonstrate that the biosensor responds specifically to copper and is not activated by zinc under the conditions tested.

\subsection{Validation} \label{validation}
In this section, the potential of a simulation platform for characterizing the response of the yeast receiver to copper stimuli, in a manner aligned with the experiments in Section~\ref{results abundance} and Section~\ref{results mRNA}, is demonstrated. Such a tool can complement experimental procedures by informing the design of subsequent trials and by providing mechanistic insight into the underlying biological processes. This is particularly important in settings where laboratory cost constitutes a limiting factor.

\begin{figure}[!hbt]
\centering
\includegraphics[width=3in]{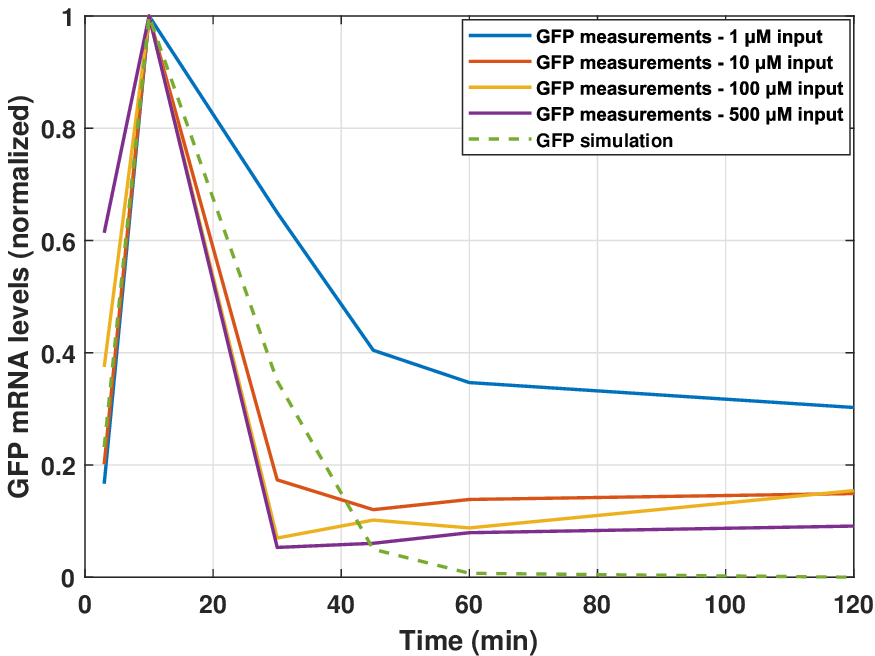}
\caption{Comparison between theoretical and experimental results across four different copper inputs. mRNA levels are reported.}
\label{validation mRNA}
\end{figure}

\begin{figure}[!hbt]
\centering
\includegraphics[width=3in]{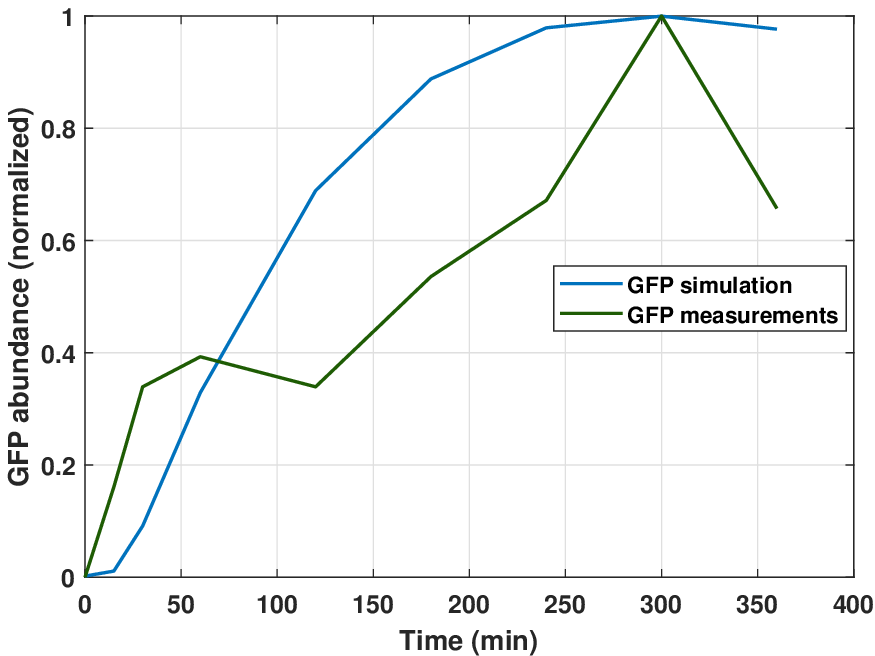}
\caption{Comparison between theoretical and experimental results for a single copper input. Protein abundance levels are reported.}
\label{validation protein}
\end{figure}

To this end, a scenario involving a single copper stimulation is considered as the input to the mathematical model of the copper detoxification pathway described in Section~\ref{system model} and specified further in Appendix~\ref{rx}. The GFP reporter is treated as the model output, and theoretical mRNA transcript and protein abundance levels are quantified as a time-evolution profile. Model versatility is examined by juxtaposing theoretical predictions under 1,10,100 and 500~$\mu$M of copper input, with the GFP transcript measurements reported in Section~\ref{results mRNA} at corresponding time points. For protein abundance levels reported in Section~\ref{results abundance}, GFP predictions are limited to a single copper input of 100~$\mathrm{\mu M}$. The resulting validation comparisons are presented in Fig.~\ref{validation mRNA} and Fig.~\ref{validation protein} respectively.

To facilitate comparison of the time-evolution profiles, measured and simulated results were normalized to their respective maximum values. Differences in absolute amplitude between theoretical and experimental traces are commonly attributed to discrepancies in quantification procedures, since the computational methods embedded in laboratory instrumentation are not readily replicated through simplified mathematical representations.

As shown in Fig.~\ref{validation mRNA}, satisfactory concordance is observed between the computed and measured profiles overall. RNA expression peaks at approximately 10 minutes and is followed by rapid degradation.
 
The observed disparities are primarily associated with the initial transcript levels and the tail end of the responses. For copper inputs below 100~$\mathrm{\mu M}$ (blue and red curves), comparatively low activation is observed at 3 minutes, resulting in closer agreement with the model predictions. In contrast, higher mRNA levels are reached within the first 3 minutes for the measured profiles that correspond to 100 and 500~$\mathrm{\mu M}$ (yellow and purple curves), whereas lower mRNA abundance is predicted by the model during this early phase. In addition, lingering transcripts are observed beyond 45 minutes in the experimental data, which are not captured by the current formulation. To quantify the observed differences, the Root Mean Square Error (RMSE) between the computed and measured profiles is reported in Table~\ref{RMSE}, where $E^{mRNA}_1,E^{mRNA}_{10},E^{mRNA}_{100}$ and $E^{mRNA}_{500}$ correspond to the errors for each concentration. Albeit the reported deviations, the agreement in peak timing, together with the overall similarity in dynamics, supports the value of the proposed model as a simulation platform for describing mRNA expression, while acknowledging its limitations.
\begin{table}[!t]
\caption{RMSE between model predictions and mRNA measurements for the GFP reporter.}
\label{RMSE}
\centering
\setlength\tabcolsep{1.5pt}
\begin{tabular}{|c|c|c|c|}
\hline
$E^{mRNA}_1$&$E^{mRNA}_{10}$&$E^{mRNA}_{100}$&$E^{mRNA}_{500}$\\ 
\hline
0.27&0.11&0.15&0.2\\
\hline
\end{tabular}
\end{table}

Notable discrepancies are also observed between the theoretical and experimental time-course protein profiles in Fig.~\ref{validation protein}. Although the peak time coincides in both cases, the predicted GFP trajectory exhibits a smooth, gradual increase toward the maximum at $t=300$ minutes. In contrast, the experimental recordings show a steeper rise with noticeable fluctuations, followed by a comparatively rapid decline in fluorescence. Such a fast decay is not captured by the current model, for which slower degradation kinetics are predicted. The kinetic parameters adopted in the GFP model are mentioned in detail in Appendix~\ref{ODEs}. It ought to be noted however, that the de-stabilized form of the sfGFP used for the experiments was not possible to be reflected in the model parameters. As a result, a wider pulse is obtained in the computed profile; nevertheless, the overall trend remains consistent, with maximum abundance reached at approximately the same time as in the measurements. The mismatch observed in Fig.~\ref{validation protein} is again quantified through the RMSE, which is found to be 0.24.

The observed differences in both graphs may be attributed to biological nuances that are not represented by the mathematical model, since emphasis has been placed on processes deemed the most relevant. Additional regulatory effects and system-specific complexities pertaining to gene expression, protein folding and stability have been neglected. It is emphasized that the presented model is intended to serve as a basis for further refinement, and is able to characterize practical response profiles of the yeast receiver to a satisfying extent.
 
\subsection{Bioluminescence Computations} \label{results luminescence}
Here, we present results obtained from an extended version of the model used for validation, which also includes its current response to copper-based stimulation. Constructing a fully developed model tailored to yeast biosensors is crucial for creating a framework that supports the incorporation of yeast MC receivers into applications in the IoBNT field.

The models describing the Lux reporter are integrated with the mathematical framework previously established for the copper detoxification pathway, and the resulting comprehensive model is employed to quantify the photocurrent response of the biosensor. The majority of the model parameters, as well as the initial concentrations of the chemical species considered in the simulations, were compiled from multiple literature sources, primarily studies on \textit{Vibrio harveyi} and \textit{Photorhabdus luminescens}. Additional kinetic investigations are necessary to obtain more accurate, yeast-specific parameter values; nevertheless, the full model presented here provides a foundational framework for subsequent refinement and optimization.

\begin{figure}[!hbt]
\centering
\includegraphics[width=3in]{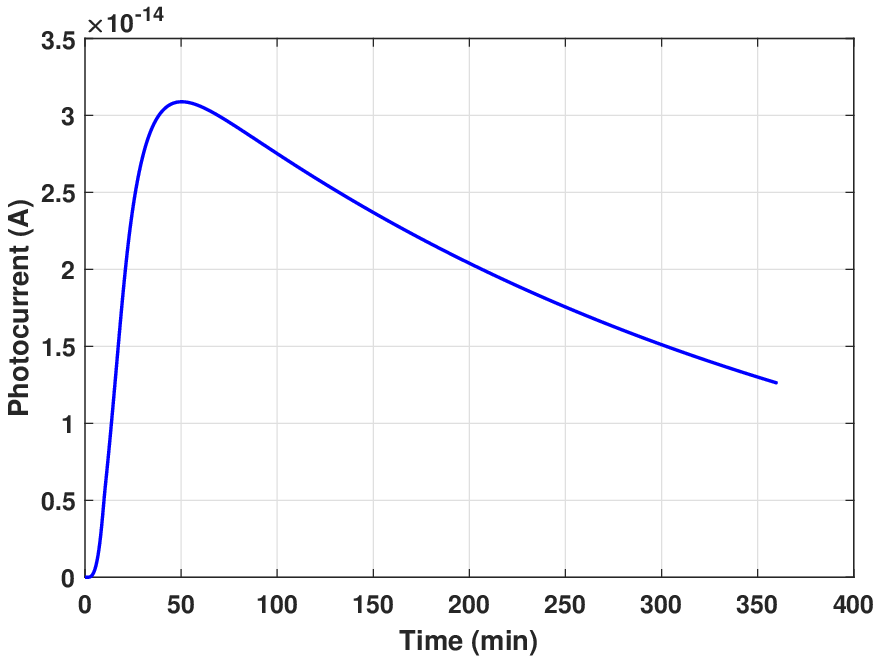}
\caption{Photocurrent response for 100 $\mu$M copper stimulation.}
\label{current profile}
\end{figure}

The copper input mobilizes the chemical reactions within the copper detoxification pathway, which culminate in the upregulation of the \textit{luxAB} gene, which is translated to the Lux enzyme. The latter triggers light emission. As in Section \ref{results abundance}, a single stimulation is considered at the receiver, where 100 $\mu$M of copper is directly supplied without the consideration of molecular propagation nuances. The photocurrent profile is shown in Fig. \ref{current profile}.

Although experimental data are not available for validation of the Lux reporter considered here, the response shown in Fig.~\ref{current profile} provides some useful qualitative insights. The time evolution of the output exhibits a peak at approximately 49 minutes and maintains a bell-shaped pulse profile with long sustain, due to the long half-life of the luciferase in \textit{Photorhabdus luminescens} \cite{Tinikul2012}. The p+/n-well/psub CMOS photodiode considered in the computations has an active diffusion area of $40 \times 40~\mu\mathrm{m}^2$ and has the highest responsivity near $\lambda=490$~nm among the other designs reported in \cite{Koklu2013}. The dark current can be estimated from the measured dark-voltage variation as follows:
\begin{equation}\label{dark current} 
I_{\mathrm{dark}} = C_{\mathrm{pd}}\,\frac{dV_{\mathrm{dark}}}{dt},
\end{equation}
where $C_{\mathrm{pd}}=1352.7$~[fF] and $\frac{dV_{\mathrm{dark}}}{dt}=18$~[mV/s]. The resulting dark current is about $24.3$~[fA], while the predicted maximum photocurrent is approximately $31$~[fA]. Therefore, the predicted photocurrent is approximately $1.3$ times larger than the estimated dark current. This comparison provides an initial indication that the photocurrent may be distinguishable from the detector baseline under ideal conditions such as those considered here. However, a complete feasibility analysis should also consider dark-current shot noise, thermal noise, detector bandwidth, integration time, and optical collection and propagation losses, which are expected to lead to a notably weaker signal and a higher noise background. Stochastic effects associated with molecular propagation and transcription-factor binding may introduce additional variability. Furthermore, the present single-cell system should be extended to account for population-level emission in a bulk cell culture medium, which may lead to a more robust photocurrent response overall.

The faster response peak compared to the GFP measurements reported in Section~\ref{results abundance} is attributed primarily to the maturation time required for GFP to reach its fluorescent form, as well as to its relatively long lifetime, which can result in lingering fluorescent molecules within the cell. In contrast, luciferase reporters are typically characterized by minimal maturation time \cite{MazoVargas2014,Baldwin1995}, therefore enabling more prompt protein-level responses and a shorter-lived output. Such effects have been incorporated into the model in order for the corresponding physical behavior to be reflected.

To further assess the proposed model, a sensitivity analysis is conducted in order to quantify changes in the biosensor response under parameter perturbations. In this manner, robustness of the predicted dynamics to specific parameter variations is examined, while response trends under alternative parameter sets can also be explored. Such an analysis can be useful for identifying dominant mechanisms and, in principle, for guiding the design of modified biosensors that incorporate 'mutant' receiver configurations.

Several parameters from the equations in Appendix~\ref{ODEs} were varied, including the maximum copper uptake rate constant (Eq.~(\ref{uptake rate})), the Ace1--DNA promoter dissociation constant (Eq.~(\ref{transcription rate})), the \textit{luxAB} mRNA degradation rate (Eq.~(\ref{LuxmRNA})), the LuxC forward catalytic rate (Eq.~(\ref{LuxC rate})), the LuxAB enzyme degradation rate (Eq.~(\ref{Luxab deg})), and the hydroxyflavin de-excitation rate (Eq.~(\ref{Dlux})). In each case, only one parameter was varied while all remaining parameters were kept fixed. As in the previous simulations, a single copper stimulation of 100~$\mu$M is considered and the resulting time-course response is evaluated across the tested parameter values.

\begin{figure}[htpb]
  \centering
  \subfloat[Maximum copper uptake rate constant sweep.]{\includegraphics[width=1.65in]{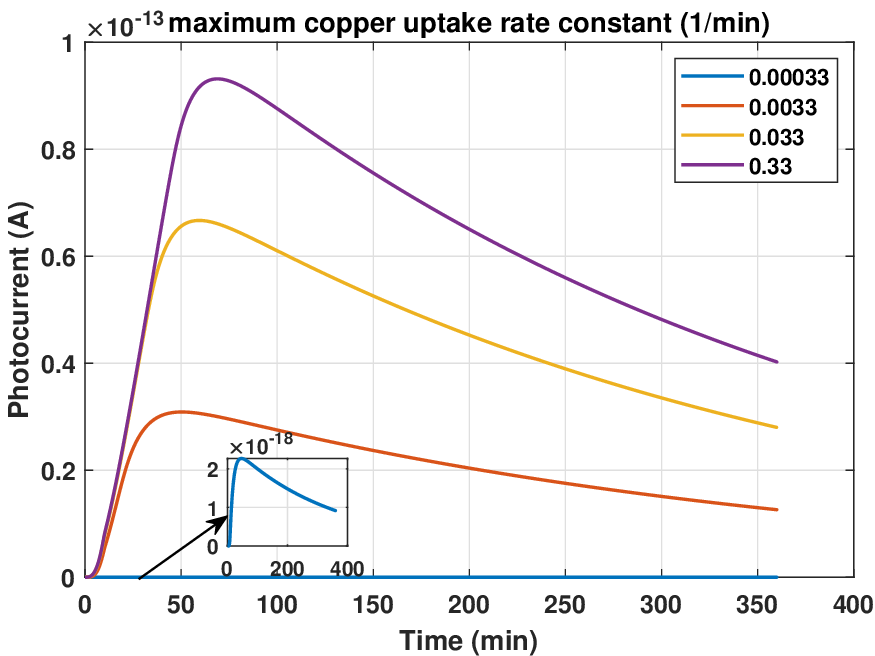}%
 \label{kupt}}
  \hfil
   \subfloat[Dissociation constant for the binding of Ace1 to the \textit{CUP1} DNA promoter sweep.]{\includegraphics[width=1.7in]{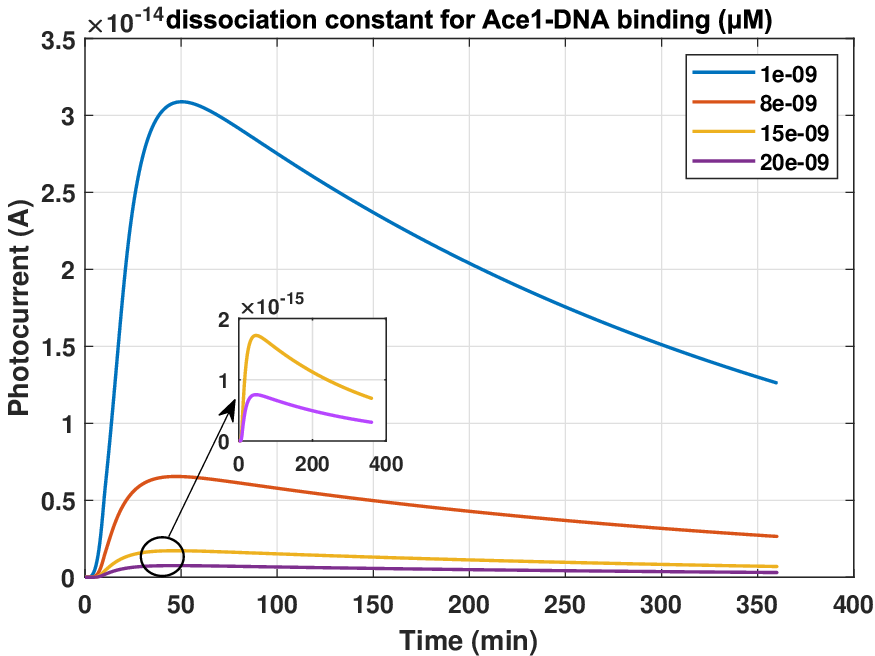}%
 \label{KDmRNA}}
  \hfil
     \subfloat[\textit{luxAB} mRNA degradation rate sweep.]{\includegraphics[width=1.7in]{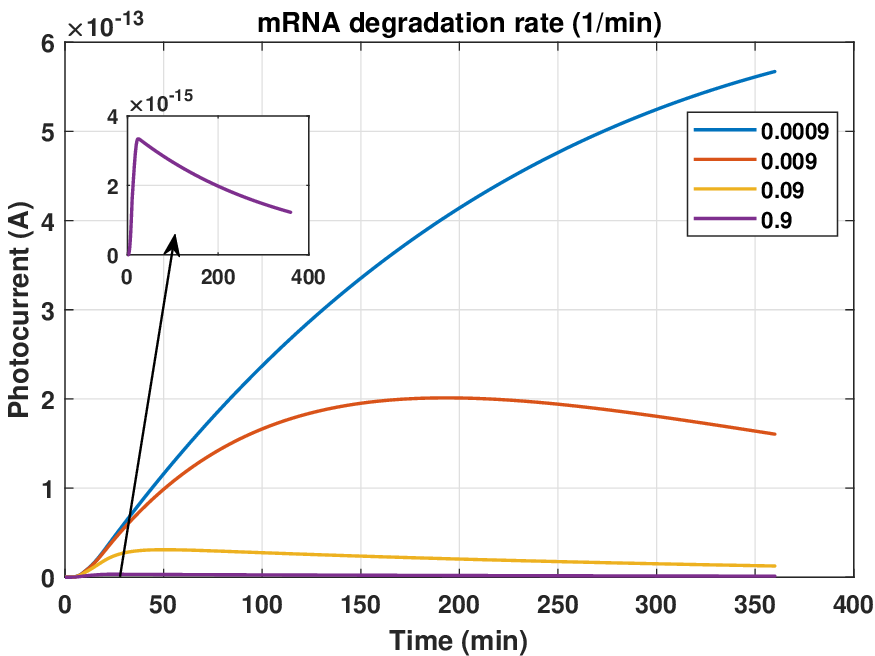}%
 \label{kdegmRNA}}  
  \hfil
    \subfloat[Forward catalysis rate of LuxC sweep.]{\includegraphics[width=1.7in]{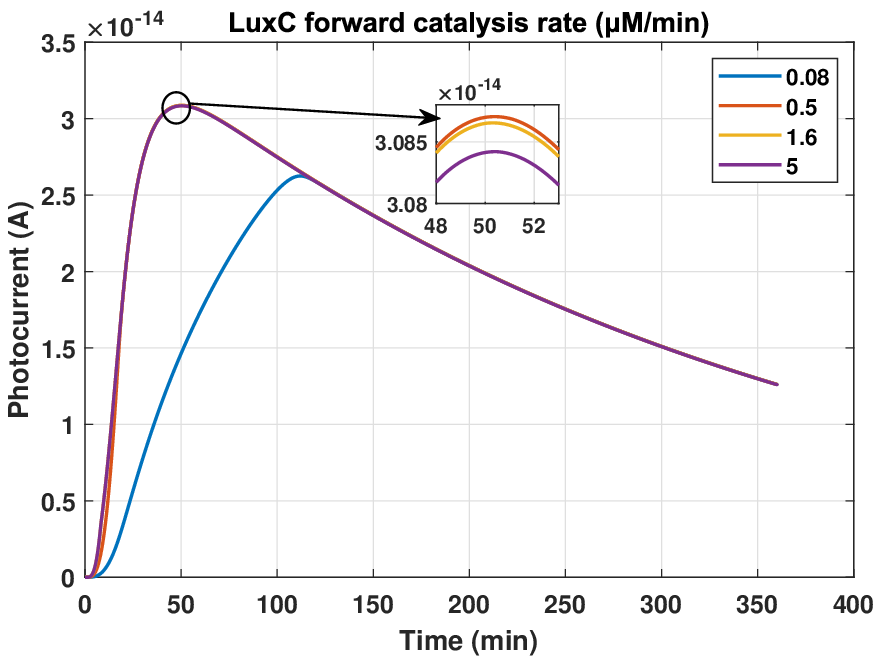}%
 \label{VLuxC}}
  \hfil
    \subfloat[LuxAB protein abundance degradation rate sweep.]{\includegraphics[width=1.7in]{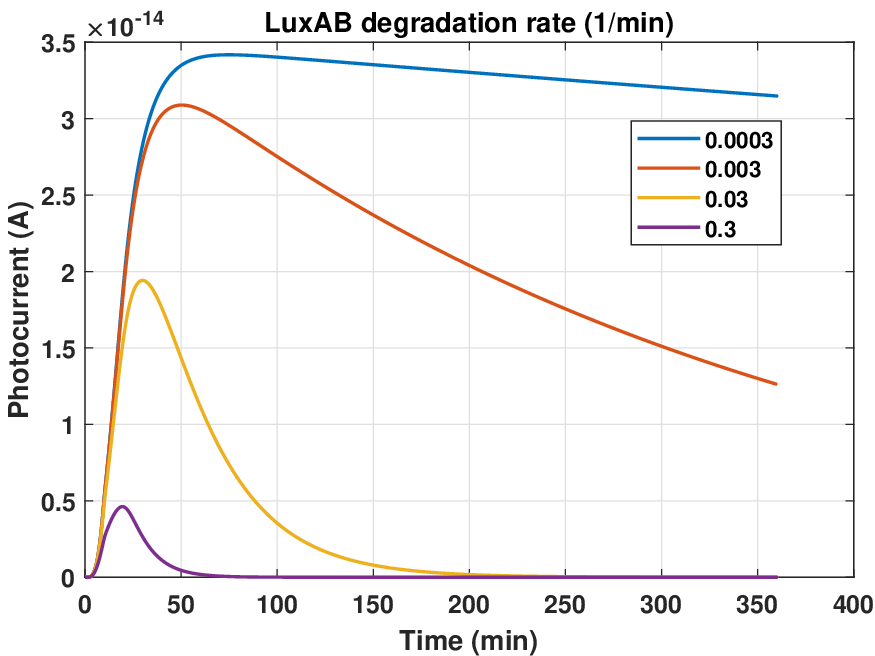}%
 \label{kdeglux}}
  \hfil
      \subfloat[De-excitation and de-hydration rate sweep of hydroxyflavin, after light emission.]{\includegraphics[width=1.7in]{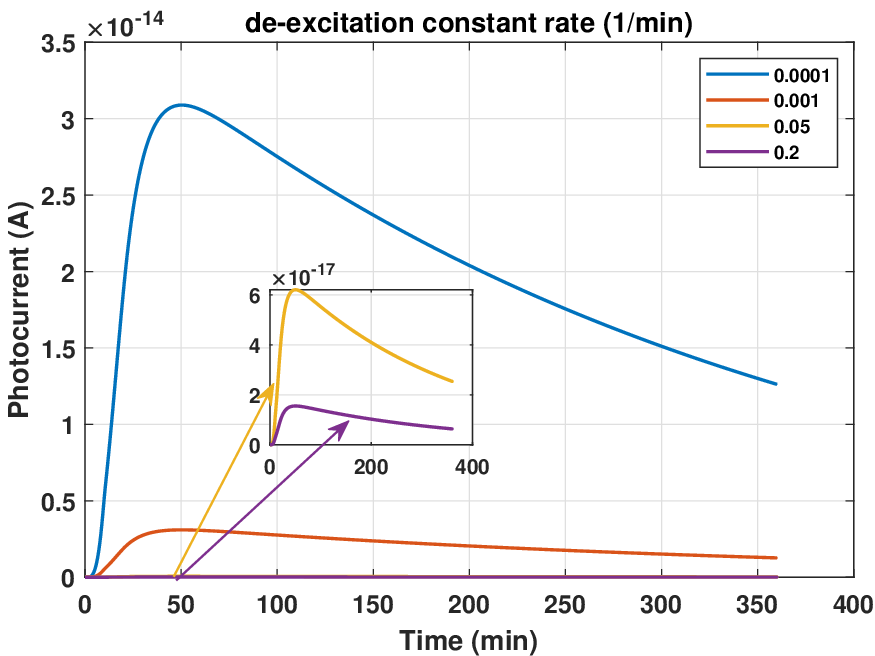}%
 \label{k4}}
  \hfil
  \caption{Parameter sweeps for 6 model parameters.}
  \label{param sweep}
\end{figure}

Inspection of the plots in Fig.~\ref{param sweep} indicates that parameter variations primarily affect the amplitude and peak time, depending on the parameter being perturbed, while pulse width was less altered. As shown in Fig.~\ref{kupt}, larger values of the maximum copper uptake constant result in increased current amplitude, albeit at the expense of a delayed peak and slightly larger pulse width. This behavior can be attributed to enhanced copper influx over the course of the simulated experiment, which prolongs the buildup of downstream species in the reaction chain. 

A less pronounced increase in amplitude is observed in Fig.~\ref{VLuxC}, where increasing the forward catalytic rate of LuxC produces a slightly stronger current response and a markedly earlier peak. However, the peak amplitude approaches a plateau at higher catalytic rates, suggesting that LuxC is no longer the rate-limiting step and that other processes within the Lux pathway limit the response. The higher current output observed for values of $V^+_{LuxC}$ above $0.08 \mu\mathrm{M\,min^{-1}}$ can be attributed to increased aldehyde production, which increases substrate availability for LuxAB-driven light emission.

In contrast, in Fig.~\ref{KDmRNA} and Fig.~\ref{k4}, parameter perturbations introduce stronger variation in pulse amplitudes. Reduced dissociation constant values for Ace1 binding to DNA promoters correspond to stronger binding affinity, thereby increasing transcript production in the nucleus and ultimately increasing LuxAB abundance and photon generation directly. Notable increase of the dissociation constant may result in faster time peaks by a very small margin, albeit at the expense of a dramatic decrease of the output current. The pulse width of the response remains largely unaffected. Altering the de-excitation rate in Fig.~\ref{k4} also results in much varied responses inferring high response sensitivity: faster de-excitation of the hydroxyflavin intermediate (see Fig.~\ref{light cycle}) reduces photon yield, whereas smaller values lead to brighter emission and to a severely increased photodetector response, without influencing peak time and pulse width.

The influence of Lux degradation rates on the current output is examined in Fig.~\ref{kdegmRNA} and Fig.~\ref{kdeglux} for luciferase mRNA and protein, respectively. In both cases, higher degradation rates yield diminished responses with narrower pulse widths, whereas lower degradation rates produce more intense and long-lasting responses. Current amplitude changes are more pronounced for mRNA degradation variations, while pulse width is markedly altered at higher protein degradation values. Peak instance is similarly affected in both cases. This suggests that incorporation of protein-degradation tags, through which accelerated turnover can be introduced, can enable responses that more closely track promoter activity and partially counteract translation delay. However, a tradeoff is introduced between detection rate and detection reliability in an MC system, since faster but weaker output pulses are more susceptible to noise and losses, thereby increasing the probability of erroneous detection.


The effects discussed above are summarized in Table~\ref{metrics}, where the current pulse metrics (amplitude, peak time, and pulse width), are provided for every plot of Fig. \ref{param sweep}. Pulse width is defined here as the time interval during which the signal remains above of the 50\% of its maximum current amplitude. If the signal does not fall below the 50\% threshold within the simulated time window, the pulse end time is set equal to the final simulation time point. Moreover, the corresponding variances are reported in Table \ref{variances}.

\begin{table}[!t]
\caption{Amplitude, Peak Time and Pulse Width values (parameter sweep).}
\label{metrics}
\centering
\setlength\tabcolsep{1.5pt}
\begin{tabular}{|p{1.3cm}|p{1.1cm}|p{1.3cm}|p{1.45cm}|c|c|p{0.75cm}|}
\hline
parameter&$k_{upt,max}$&$K_{DmRNA}$&$k_{deg,mRNA}$&$V_{LuxC}$&$k_{deg,LuxAB}$&$k_4$\\
\hline
Peak1 [fA] (blue)&0.002&30.1&567.2&25.61&34.2&31\\
\hline
Peak2 [fA] (orange)&31&6.55&201.14&30.14&31&3.1\\
\hline
Peak3 [fA] (yellow)&66.7&1.72&31&31&19.42&0.062\\
\hline
Peak4 [fA] (purple)&93.15&0.76&3.34&30.12&4.62&0.016\\
\hline
PeakT1 [min] (blue)&46.73&50.41&360&112.3&75&50.41\\
\hline
PeakT2 [min] (orange)&50.4&47.4&193.23&50.23&50.41&50.4\\
\hline
PeakT3 [min] (yellow)&59.44&47&50.41&50.2&29.92&50.33\\
\hline
PeakT4 [min] (purple)&69&47&23.91&50.2&19.5&50.3\\
\hline
Width1 [min] (blue)&275&275.2&-&301.75&341.5&275.2\\
\hline
Width2 [min] (orange)&275.2&275&308.95&275&275.2&275.2\\
\hline
Width3 [min] (yellow)&277.3&275&275.18&275.1&50.73&275.2\\
\hline
Width4 [min] (purple)&280.65&275&245.65&275.14&22.11&275.14\\
\hline
\end{tabular}
\end{table}

\begin{table}[!t]
\caption{Amplitude, Peak Time and Pulse Width variances (\%, parameter sweep).}
\label{variances}
\centering
\setlength\tabcolsep{1.5pt}
\begin{tabular}{|c|c|c|c|}
\hline
parameter&PeakVar&PeakTimeVar&WidthVar\\ 
\hline
$k_{upt,max}$&100&32.25&2.1\\
\hline
$K_{DmRNA}$&97.6&7&0.1\\
\hline
$k_{deg,mRNA}$&99.42&93.36&23.1\\
\hline
$V_{LuxC}$&15.1&55.21&8.94\\
\hline
$k_{deg,LuxAB}$&86.5&74&93.52\\
\hline
$k_4$&100&0.21&0.01\\
\hline
\end{tabular}
\end{table}

\begin{figure}[htpb]
  \centering
  \subfloat[Ctrs initial concentration sweep.]{\includegraphics[width=1.7in]{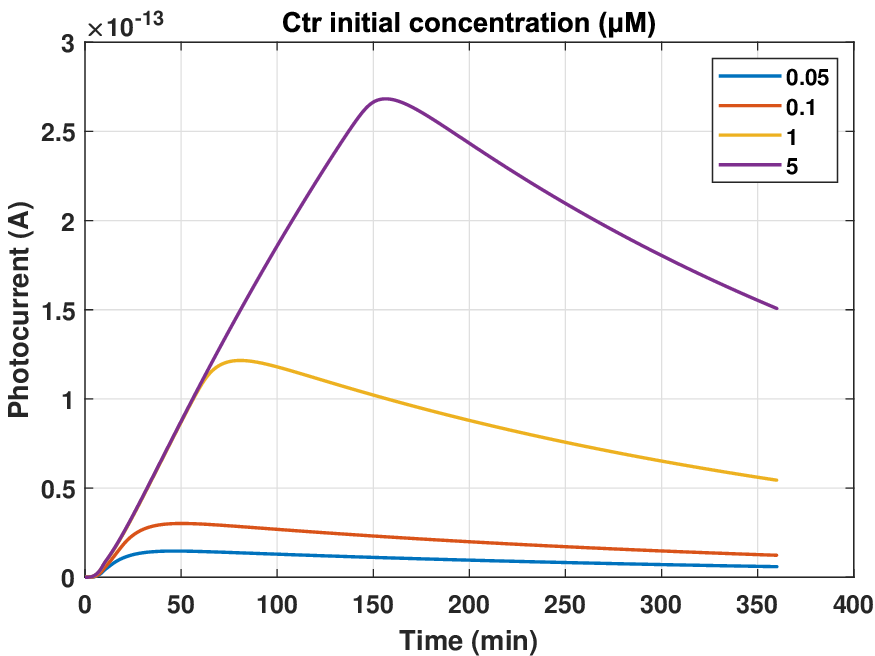}%
 \label{Ctr1 sweep}}
  \hfil
  \subfloat[FMN initial concentration sweep.]{\includegraphics[width=1.7in]{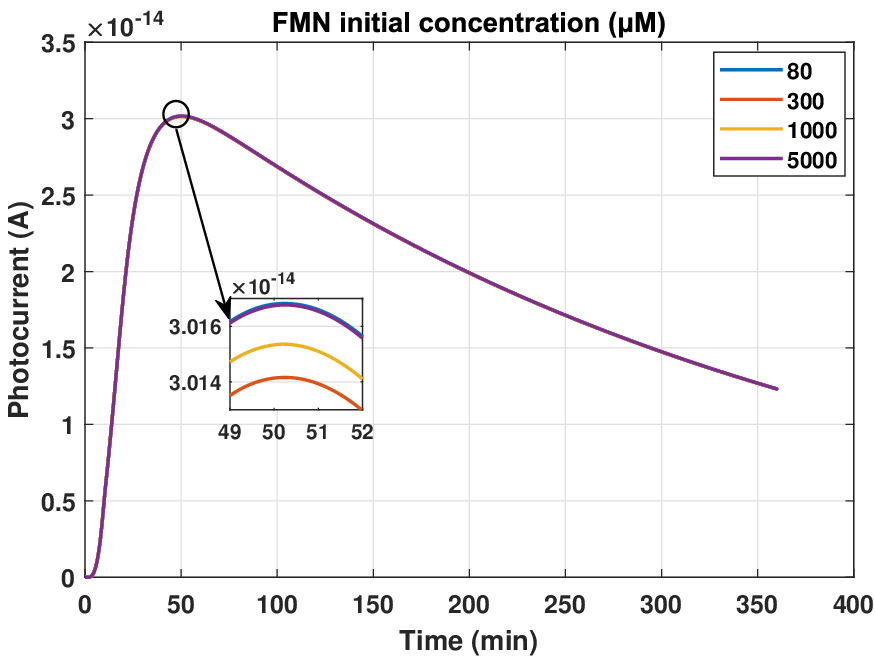}%
 \label{FMN sweep}}
  \hfil
    \subfloat[RCOOH initial concentration sweep.]{\includegraphics[width=1.7in]{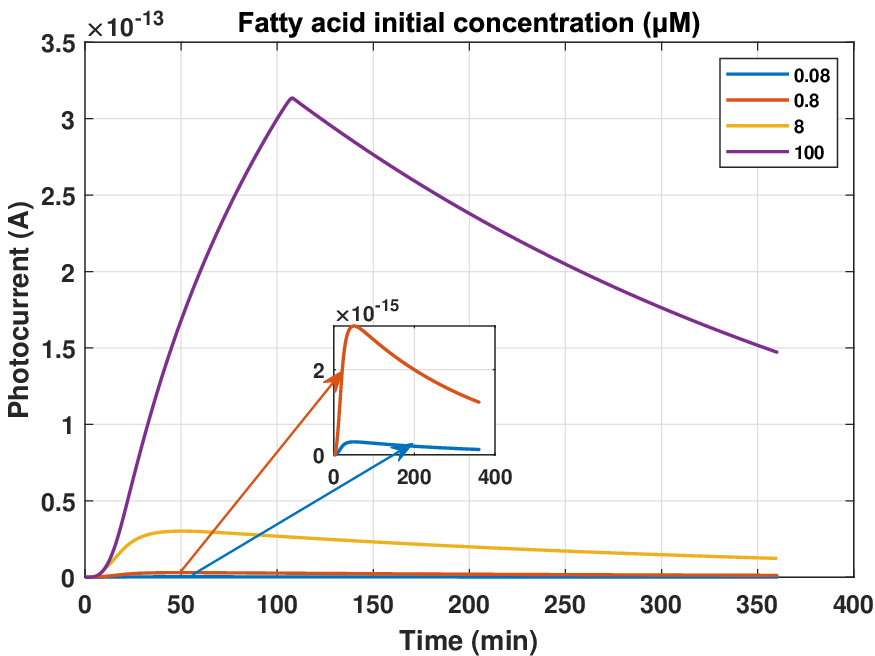}%
 \label{Aldehyde sweep}}
  \hfil
    \subfloat[Cu dose-response curves for the output photocurrent.]{\includegraphics[width=1.7in]{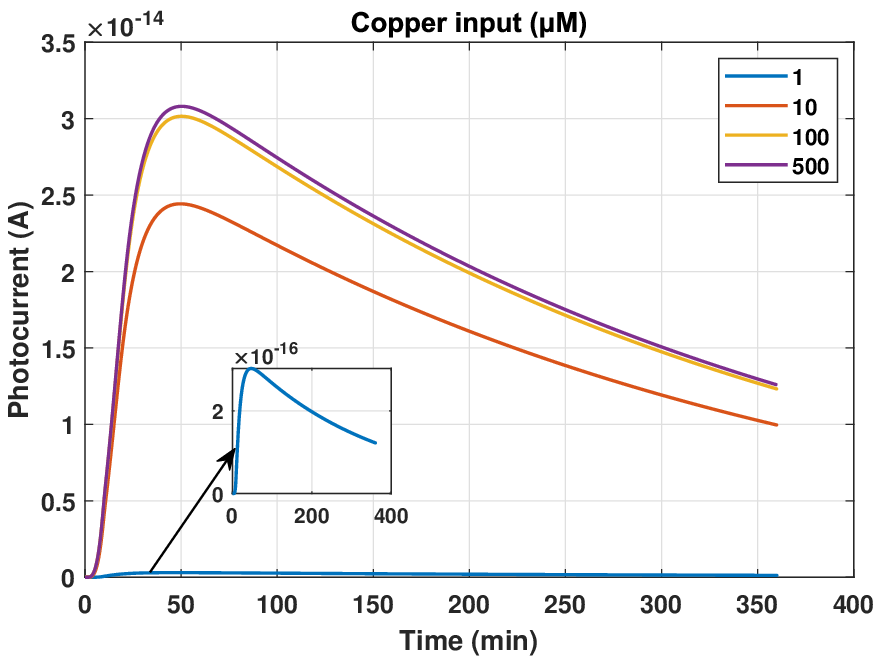}%
 \label{copper sweep}}
  \hfil
  \caption{Different initial concentrations for Ctr transporters, FMN and fatty acids. A copper dose-response is also included.}
  \label{concentrations sweep}
\end{figure}

To complement the sensitivity analysis with respect to internal model parameters, the influence of the initial amounts of three chemical compounds is additionally examined, namely the transporters Ctr1/3, the cofactor FMN, and intracellular fatty acids, as shown in Fig.~\ref{concentrations sweep}. In addition, a simulated dose-response graph is provided, where the output current is reported for different copper input levels.

Pronounced sensitivity is observed with respect to the initial concentration of the Ctr transporters. As shown in Fig.~\ref{Ctr1 sweep}, increased transporter abundance results in a substantial increase in current amplitude. The output pulse also widens as the Ctr concentration increases up to $1~\mu\mathrm{M}$, while the current peak is progressively delayed. In contrast, the output current remains largely unaffected by variations in the initial FMN concentration, as shown in Fig.~\ref{FMN sweep}. This suggests that FMN is not a limiting substrate within the tested concentration range and may already be present at effectively saturating levels.

In Fig.~\ref{Aldehyde sweep}, low initial fatty-acid concentrations, such as $0.08~\mu\mathrm{M}$, substantially attenuate the current output without noticeably affecting the pulse width. Higher fatty-acid concentrations support increased aldehyde production and, consequently, greater substrate availability for LuxAB-driven light emission, resulting in stronger current responses. A sharp increase in current amplitude is observed when the RCOOH concentration is increased from $8$ to $100~\mu\mathrm{M}$. However, the highest concentration also introduces a delay in the current peak and a modest increase in pulse width. Finally, the simulated copper dose-response curves in Fig.~\ref{copper sweep} show that different copper input levels produce distinct photocurrent profiles that differ mainly in amplitude. Changes in peak time and pulse width remain comparatively limited, while the current amplitude approaches saturation at the higher copper concentrations.

In addition to the profiles presented in Fig.~\ref{concentrations sweep}, the current-pulse metrics are again summarized in Table~\ref{metrics2}, while the corresponding variances are reported in Table~\ref{variances2}.

\begin{table}[!t]
\caption{Amplitude, Peak Time and Pulse Width values (initial concentration sweep).}
\label{metrics2}
\centering
\setlength\tabcolsep{1.5pt}
\begin{tabular}{|c|c|c|c|c|}
\hline
chemical compound&Ctr1/3&FMN&RCOOH&Cu input\\
\hline
Peak1 [fA] (blue)&15&30.2&0.3&0.3\\
\hline
Peak2 [fA] (orange)&30.2&30.2&3.02&24.43\\
\hline
Peak3 [fA] (yellow)&122&30.2&30.16&30.2\\
\hline
Peak4 [fA] (purple)&268.3&30.2&313.5&30.81\\
\hline
PeakT1 [min] (blue)&47&50.23&50.22&47.16\\
\hline
PeakT2 [min] (orange)&50.23&50.26&50.22&49.6\\
\hline
PeakT3 [min] (yellow)&80.83&50.2&50.22&50.23\\
\hline
PeakT4 [min](purple)&156.5&50.26&108&50.3\\
\hline
Width1 [min] (blue)&274.92&275.1&275.14&275\\
\hline
Width2 [min] (orange)&275.14&275.12&275.14&275.03\\
\hline
Width3 [min] (yellow)&285.6&275.1&275.14&275.14\\
\hline
Width4 [min] (purple)&287.16&275.1&292.5&275.14\\
\hline
\end{tabular}
\end{table}

\begin{table}[!t]
\caption{Amplitude, Peak Time and Pulse Width variances (\%, initial concentration sweep).}
\label{variances2}
\centering
\setlength\tabcolsep{1.5pt}
\begin{tabular}{|c|c|c|c|}
\hline
chemical compound&PeakVar&PeakTimeVar&WidthVar\\ 
\hline
Ctr1/3&94.53&70.23&4.26\\
\hline
FMN&0.09&0.14&0.01\\
\hline
RCOOH&99.9&53.4&5.92\\
\hline
Cu&99&6.2&0.1\\
\hline
\end{tabular}
\end{table}

\begin{figure}[!hbt]
\centering
\includegraphics[width=3in]{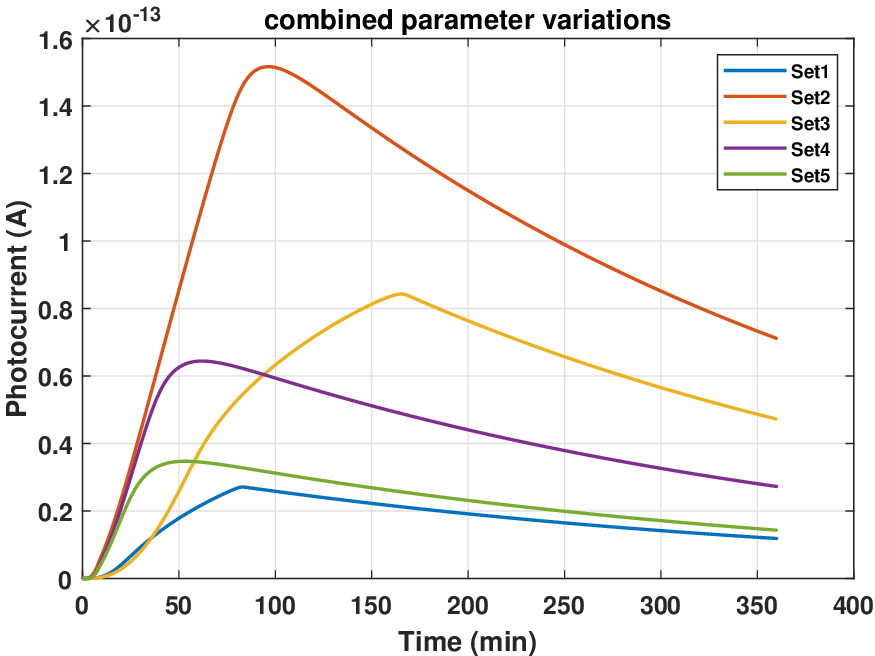}
\caption{Combined parameter sweep.}
\label{mixed combinations}
\end{figure}

Although individual parameter perturbations provide useful insight into the influence of key model parameters on the output current, evaluating the biosensor under simultaneous parameter changes is also important. Since the model includes several parameters obtained from non-yeast systems, varying multiple parameters together can provide a broader representation of parameter uncertainty. For this purpose, the selected parameters include values that were either obtained from non-yeast systems or assumed due to the lack of available data. These parameters are the LuxE forward catalytic rate ($V^+_{LuxE}$, Eq.~(\ref{LuxE rate})), the FMN dissociation constant ($K{D,F}$, Eq.~(\ref{Dfrp})), the oxygen reaction rate constant ($k_O$, Eq.~(\ref{Dlux})), the Michaelis constant of ATP ($K_{m,ATP}$, Eq.~(\ref{D_RCOOH})), the dissociation constant of fatty acids ($K_{D,RCOOH}$, Eq.~(\ref{D_RCOOH})), the Ctr trimerization rate ($k_{b1}$, Eq.~(\ref{trimerization})), and the binding rate constant of the substrate $\mathrm{FMNH_2}$ ($k_1$, Eq.~(\ref{Dlux})). The corresponding parameter sets are provided in Table~\ref{parameter sets}. The results shown in Fig.~\ref{mixed combinations} provide further insight into how combinations of parameter changes affect the photocurrent profile.

Out of the parameters tested, the binding rate $k_{b1}$ appears to exert the strongest influence on the output current. Higher values of this parameter across the different sets, consistently resulted in increased amplitude. This result is consistent with those shown in Fig.~\ref{kupt} and Fig.~\ref{Ctr1 sweep}, indicating that copper transport through the Ctr proteins is among the strongest determinants of the current amplitude. The combination of higher $V^+_{LuxE}$ and $k_1$ also stands out, as it is associated with a steeper increase in photocurrent during the rising phase in sets 2, 4, and 5. In contrast, lower values of these two rates are generally associated with more gradual responses, as observed in sets 1 and 3. The suggested influence of $V^+_{LuxE}$ is also consistent with the results shown in Fig.~\ref{VLuxC}, which indicate that the forward catalytic rates of both enzymes involved in aldehyde production affect the slope of the response.

It is imperative to note that, the temporal characteristics of the biosensor's response are expected to play a pivotal role in the achievable data rate of the envisioned communication system. When viewed from an MC perspective, lower response latency combined with smaller output pulse width will contribute to real-time, fast detection of molecular events. In case of re-stimulation capability of the system under consideration, wider pulses may cause overlap between output pulses as observed in previous work \cite{Ntetsikas2024}, undermining detection performance.

\begin{table}[!t]
\caption{Parameter sets used for a mixed parameter sweep.}
\label{parameter sets}
\centering
\setlength\tabcolsep{1.5pt}
\begin{tabular}{|c|c|c|c|c|c|}
\hline
Parameter&Set1&Set2&Set3&Set4&Set5\\ 
\hline
$V^+_{LuxE}$ [$\mu M\,min^{-1}$]&0.1&10&0.05&1.2&100\\
\hline
$K_{D,F}$ [$\mu M$]&0.0017&1.7&0.17&0.017&$1.7\cdot 10^{-4}$\\
\hline
$k_O$ [$min^{-1}$]&210000&21000&210&2100&700\\
\hline
$K_{m,ATP}$ [$\mu M$]&0.002&0.02&0.2&$2\cdot 10^{-4}$&$2\cdot 10^{-5}$\\
\hline
$K_{D,RCOOH}$ [$\mu M$]&0.01&1&0.1&$10^{-3}$&$10^{-4}$\\
\hline
$k_{b1}$ [$\mu M^{-2}\,min^{-1}$]&0.1&1000&100&10&1\\
\hline
$k_1$ [$\mu M^{-1}\,min^{-1}$]&1020&102&10.2&10200&102000\\
\hline
\end{tabular}
\end{table}

Overall, the proposed mathematical framework is able to adapt to parameter variations that reflect different experimental conditions to a satisfactory extent. This is crucial, due to the scarcity of reported values for yeast systems in the literature. Since the model includes a large number of parameters that couple the copper detoxification pathway to the Lux reporter—many of which have not been quantitatively determined for eukaryotic systems—further experimentation, together with refinement of modeling assumptions, is expected to yield a more nuanced theoretical tool. In this manner, a digital-twin representation of a yeast-derived biosensor can be progressively approached.


\section{Conclusion}\label{Conclusion}
In this work, \textit{S. cerevisiae} is leveraged for the development of an MC receiver capable of reporting copper presence in its vicinity. A theoretical framework is presented for characterization of the dominant chemical reactions within the yeast copper detoxification pathway, while the system output is quantified through a coupled non-linear ODE model. Validation attempts of GFP model predictions against experimental trials conducted on a dedicated MC testbed, supports the proposed computational approach. Hence, it provides the basis for a simulation platform suitable for virtual experimentation.

The framework is further extended by replacing the GFP reporter with an auto-luminescent system capable of generating optical output without external light excitation, thereby enabling coupling of the yeast receiver to a photodetector unit and supporting integration into biosensing devices such as Smart Pills. The resulting model is used to characterize the photocurrent response of the proposed yeast-derived biosensor, while sensitivity analysis provides insight into key mechanisms underlying the light pathway.

Overall, this paper constitutes a step toward establishing MC as a key contributor for IoBNT applications. Future work will focus on the implementation of a new testbed that integrates the Lux system in yeast cells, and experimental validation of the Lux model extension to further strengthen its predictive value. Moreover, a transmission model coupled with the molecular propagation channel that accounts for the effects of the physical Tx-Rx separation, will be integrated together with the receiver into a comprehensive communication framework. Noise analysis pertaining to all different parts of the communication model will support error performance characterization and detection methods. Finally, investigation of the re-stimulation capabilities of the system both physically and computationally will be pursued, and achievable data rates will be assessed.

\appendices
\section{Receiver Model}\label{rx}
\subsection{Transcriptional activation function $f_{act}$}\label{fact derivation}
In this work, MRE occupancy is assumed to pertain to two distinct occupation states (OS), namely $OS_0$ when there is no TF bound and $OS_1$ when TF molecules are attached to the MREs of the promoter. The fractional transcriptional level (FTL) for a single MRE that is bound by Ace1 is given as follows:

\begin{equation}\label{OS}
FTL = \frac{OS_1}{OS_0+OS_1}
\end{equation}
Binding creates the transition from $OS_0$ to $OS_1$ and unbinding reverts back to $OS_0$. By considering rapid binding/unbinding of Ace1 to and from the MRE, and the fact that Ace1 binds to the promoter as a monomer \cite{pena1998}, it is straightforward to express the ratio $\frac{OS_0}{OS_1}$ as follows:
\begin{equation}
\frac{OS_0}{OS_1}= \frac{K_{D,mRNA}}{[Ace1Cu]}
\end{equation}
Where the dissociation constant $K_{D,mRNA}$ is defined as the ratio of the unbinding constant rate of Ace1 from the $OS_1$ conformation towards the binding rate of Ace1 to the $OS_0$ conformation.

Thus, the FTL can be written as:
\begin{equation}\label{FTL}
FTL = \frac{1}{\frac{K_{D,mRNA}}{[Ace1Cu]}+1}
\end{equation}
Moreover, Ace1 binds to four MREs \cite{pena1998}, which we assume that contribute to the gene induction in an independent manner. To reflect all possible contributions of MRE occupancy within the DNA promoter,  Eq.~\eqref{FTL} is modified accordingly to:

\begin{equation}\label{FTL2}
FTL= \frac{1}{\sum_{n=1}^{4} \bigg(\frac{K_{D,mRNA}}{[Ace1Cu]}\bigg)^n+1}
\end{equation}
As FTL scales the maximum transcriptional induction possible depending on the possible conformations available for the system under consideration, the final form of the transcriptional activation function $f_{act}$ takes the form:
\begin{equation}\label{fact}
f_{act} = k_{tr,max}\,FTL \nonumber
\end{equation}
With $k_{tr,max}$ being the maximum transcription rate constant. This approach matches the approach also followed in \cite{Mitre2016} to a large extent.

\subsection{Rate equations based on the King-Altman schematic method} \label{KA Method}
\subsubsection{Denominator '$D$' in Eq.~(\ref{EX vel1})}
\begin{figure}[htpb]
  \centering
  \subfloat[Pattern a ($E_{FRP}F$).]{\includegraphics[width=1.3in]{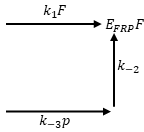}%
 \label{patt a2}}
  \hfill
  \subfloat[Pattern b ($E_{FRP}F$).]{\includegraphics[width=1.3in]{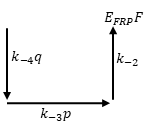}%
 \label{patt b2}}
 \hfill
   \subfloat[Pattern c ($E_{FRP}F$).]{\includegraphics[width=1.3in]{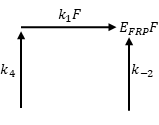}%
 \label{patt c2}}
 \hfill
    \subfloat[Pattern d ($E_{FRP}F$).]{\includegraphics[width=1.3in]{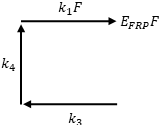}%
 \label{patt d2}}
 \hfill
  \caption{Legitimate King-Altman patterns for the flavin reduction reaction for the enzyme form $E_{FRP}F$.}
  \label{legitimate patterns2}
\end{figure}

As mentioned in Section~\ref{Bioluminescence}, the denominator D comprises all products stemming from the valid King-Altman patterns for every enzyme form in the reaction between FRP, FMN and NADPH. Since there are four legitimate patterns for every enzyme form and the patterns for EX have already been derived, we proceed by outlining the rest of the 12 diagrams that use Fig.~\ref{patterns} as a template. Irreversible product release is maintained for all diagrams.
Hence, for the enzyme form $E_{FRP}F$, the legitimate diagrams are shown in Fig.~\ref{legitimate patterns2}.

The legitimate patterns for the enzyme form $EQ$ are the following, shown in Fig.~\ref{legitimate patterns3}:
\begin{figure}[htpb]
  \centering
  \subfloat[Pattern a ($EQ$).]{\includegraphics[width=1.3in]{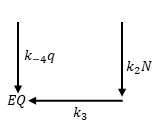}%
 \label{patt a3}}
  \hfill
  \subfloat[Pattern b ($EQ$).]{\includegraphics[width=1.3in]{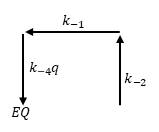}%
 \label{patt b3}}
 \hfill
   \subfloat[Pattern c ($EQ$).]{\includegraphics[width=1.3in]{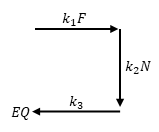}%
 \label{patt c3}}
 \hfill
    \subfloat[Pattern d ($EQ$).]{\includegraphics[width=1.3in]{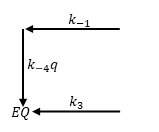}%
 \label{patt d3}}
 \hfill
  \caption{Legitimate King-Altman patterns for the flavin reduction reaction for the enzyme form $EQ$.}
  \label{legitimate patterns3}
\end{figure}

\begin{figure}[htpb]
  \centering
  \subfloat[Pattern a ($E_{FRP}$).]{\includegraphics[width=1.3in]{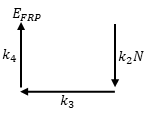}%
 \label{patt a4}}
  \hfill
  \subfloat[Pattern b ($E_{FRP}$).]{\includegraphics[width=1.3in]{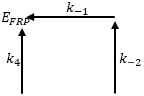}%
 \label{patt b4}}
 \hfill
   \subfloat[Pattern c ($E_{FRP}$).]{\includegraphics[width=1.4in]{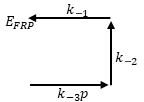}%
 \label{patt c4}}
 \hfill
    \subfloat[Pattern d ($E_{FRP}$).]{\includegraphics[width=1.1in]{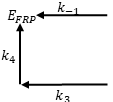}%
 \label{patt d4}}
 \hfill
  \caption{Legitimate King-Altman patterns for the flavin reduction reaction for the free enzyme form $E_{FRP}$.}
  \label{legitimate patterns4}
\end{figure}

And finally, for the free enzyme form $E_{FRP}$, the patterns are shown in Fig.~\ref{legitimate patterns4}.

Out of all the diagrams presented in Figs.~\ref{legitimate patterns2}, \ref{legitimate patterns3} and \ref{legitimate patterns4}, the valid ones that contribute to every enzyme form will only be those shown in Fig. \ref{patt c2}, \ref{patt d2}, \ref{patt c3}, \ref{patt a4}, \ref{patt b4}, \ref{patt d4}, due to the exclusion of product inhibition. Together with the pattern of Fig.~\ref{patt b}, these diagrams provide the products for the denominator $D_{FRP}=D_{E_{FRP}}+D_{E_{FRP}F}+D_{EX}+D_{EQ}$. Each '$D$' simply contains the products of the reaction rates depicted on the paths in the corresponding legitimate patterns. After some simple algebraic manipulation, the common denominator $D_{FRP}$ can be written as:
\begin{align}\label{Dfrp2}
D_{FRP}=k_4k_{-1}\,(k_{-2}+k_3)+ k_1\,k_4\,(k_{-2}+k_3)[F]&+\nonumber\\k_4\,k_3\,k_2\,[N]+ k_1\,k_2\,(k_4+k_3)\,[F]\,[N]&
\end{align}
The same procedure is followed to construct all the legitimate patterns for the enzymatic reactions involving LuxE and LuxC, and their derivations are omitted. Although the kinetic constants involved in the King-Altman formulations describe the chemical interactions that contribute to every enzyme form in more detail, using Cleland's form is often more convenient as comprehensive parameters like the Michaelis constant or dissociation factors are typically more often mentioned in the literature. Using both kinds of formulations can also aid in parameter estimation, simply by expressing the group parameters of Cleland's formulation as functions of the kinetic constants from the original King-Altman formulation, and proceed with proper algebraic manipulations \cite{cornish2013}.
\subsubsection{Rate of LuxAB catalysis} 
\begin{figure}[htpb]
  \centering
  \subfloat[Pattern a ($EZ$).]{\includegraphics[width=1.6in]{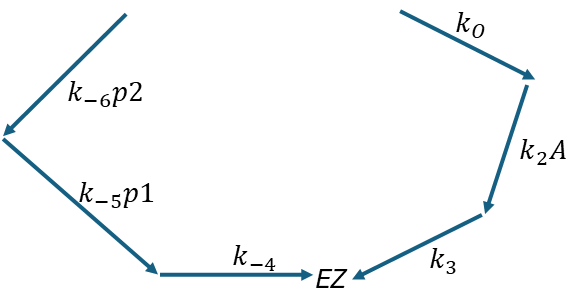}%
 \label{Lux patt a}}
  \hfill
  \subfloat[Pattern b ($EZ$).]{\includegraphics[width=1.6in]{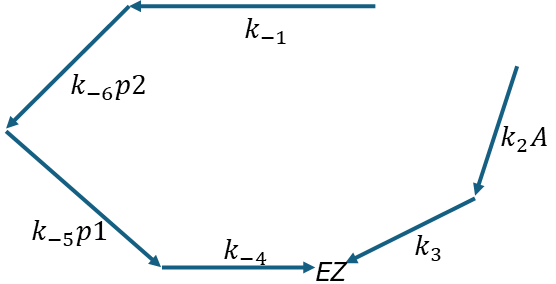}%
 \label{Lux patt b}}
 \hfill
   \subfloat[Pattern c ($EZ$).]{\includegraphics[width=1.6in]{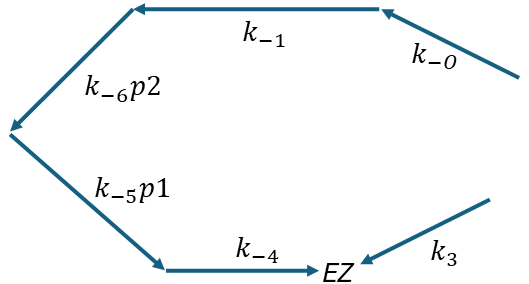}%
 \label{Lux patt c}}
 \hfill
    \subfloat[Pattern d ($EZ$).]{\includegraphics[width=1.6in]{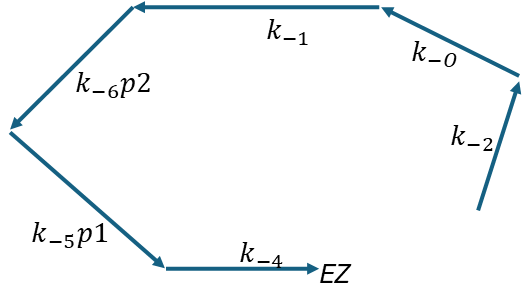}%
 \label{Lux patt d}}
 \hfill
     \subfloat[Pattern e ($EZ$).]{\includegraphics[width=1.6in]{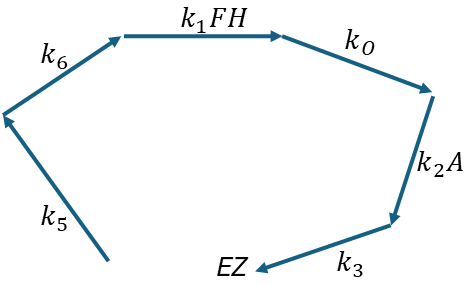}%
 \label{Lux patt e}}
 \hfill
      \subfloat[Pattern f ($EZ$).]{\includegraphics[width=1.6in]{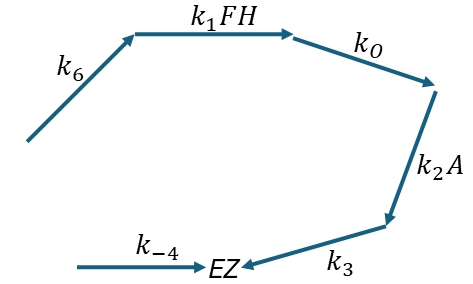}%
 \label{Lux patt f}}
 \hfill
       \subfloat[Pattern g ($EZ$).]{\includegraphics[width=1.6in]{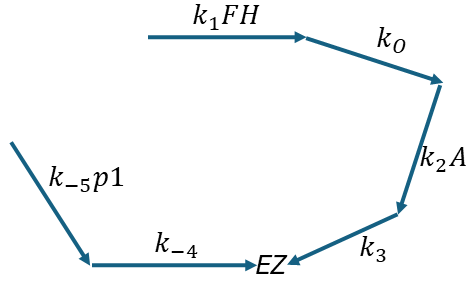}%
 \label{Lux patt g}}
 \hfill
  \caption{Legitimate King-Altman patterns for the light-emitting enzyme form $EZ$ (excited hydroxyflavin).}
  \label{legitimate patterns lux}
\end{figure}

The master pattern based on which the King-Altman patterns are drawn, is shown in Fig~\ref{lux KA}. Light emission is assumed to be generated from the enzyme form $EZ$, and the King-Altman patterns for this case are shown in Fig.~\ref{legitimate patterns lux}.

Reverse binding of products and oxygen is not considered in this cycle, nor inverse reactions of intermediates that result from chemical conformation changes. Thus, reversibility is considered only for $E_{Lux}FH$ and $EY$. Thus, the only valid pattern is the one depicted in Fig.~\ref{Lux patt e}. The concentration of $EZ$ is given as:
\begin{equation} \label{EZ}
[EZ]=\frac{[E^{Lux}_{tot}]\,N_{EZ}}{D_{Lux}}
\end{equation}
With $N_{EZ}$ given by:
\begin{equation} \label{N_EZ}
N_{EZ}=k_5\,k_6\,k_1\,k_O\,k_2\,k_3\,[FH]\,[A]
\end{equation}
The constant rate $k_2$ is defined by Eq.~(\ref{Aldehyde bind}). The derivation of $D_{Lux}$ requires all possible patterns to be drawn out for all possible enzyme forms. As this is a long and repetitive-yet straightforward procedure, we proceed by directly providing $D_{Lux}$ as follows:
\begin{align}\label{Dlux}
&D_{Lux}=k_2\,k_3\,k_4\,k_5\,k_6\,(k_O+k_{-1})[A] \nonumber\\+ &k_O\,k_1\,k_4\,k_5\,k_6\,(k_3+k_{-2})[FH] \nonumber\\+ &k_1\,k_2(k_3\,k_4\,k_5\,k_6+k_O\,k_4\,k_5\,k_6+k_O\,k_3\,k_5\,k_6+k_O\,k_3\,k_4\,k_6+\nonumber\\&k_O\,k_3\,k_4\,k_5)\,[FH]\,[A]
\end{align}
Photon emission results from Eq.~(\ref{light eq}). We considered the inclusion of key intermediate states within the light pathway to be of great importance as they can reveal important insights on the kinetic mechanisms that participate in this cycle. Also, many of the kinetic parameters were able to be adopted or inferred from the literature. 
\subsection{ODE model} \label{ODEs}
All species involved in the copper detoxification pathway are explicitly modeled. In the Lux pathway, $\mathrm{FRP}$ is treated as constant. LuxE and LuxC concentrations are incorporated into $V^+_{LuxE}$ and $V^+_{LuxC}$, and are thus treated similarly. $\mathrm{RCOOH}$, FMN and $\mathrm{NADPH}$ are modeled, yet their equations do not contain a source term. This is due to the assumption of their replenishment internally, at the end of the simulated response. Unlike those of the copper detoxification pathway, non-assumed Lux parameters are adopted from bacterial systems.

\begin{flalign}
&\frac{d[Cu]}{dt} = p0-n0 &\\
&\frac{d[Ctr1/3]}{dt} = p1-n1 &\\
&\frac{d[Ctr1/3_{trim}]}{dt} = p2-n2 &\\
&\frac{dCtr1/3_{trim}Cu}{dt} = p3-n3 &\\
&\frac{d[GSH]}{dt} = p4-n4 &\\
&\frac{d[GCu]}{dt} = p5-n5 &\\
&\frac{d[Ccs1]}{dt} = p6-n6 &\\
&\frac{d[Ccs1Cu]}{dt} = p7-n7 &\\
&\frac{d[Sod1]}{dt} = p8-n8 &\\
&\frac{d[Atx1]}{dt} = p9-n9 &\\ 
&\frac{d[Atx1Cu]}{dt} = p10-n10 &\\
&\frac{d[Ccc2]}{dt} = p11-n11 &\\
&\frac{d[Ace1]}{dt} = p12-n12 &\\
&\frac{d[Ace1Cu]}{dt} = p13-n13 &\\
&\frac{d[GFP_{mRNA}]}{dt} = p14-n14 &\\
&\frac{d[GFP_{nascent}]}{dt} = p15-n15 &\\
&\frac{d[GFP_{mature}]}{dt} = p16-n16 &\\
&\frac{d[luxAB_{mRNA}]}{dt} = pluxAB_{mRNA}-nluxAB_{mRNA} &\\
&\frac{d[LuxAB]}{dt} = pLuxAB-nLuxAB &\\
&\frac{d[FRP]}{dt} = pFRP-nFRP &\\
&\frac{d[RCOOH]}{dt} = pRCOOH-nRCOOH &\\
&\frac{d[LuxE-RCO]}{dt} = pLuxERCO-nLuxERCO&\\
&\frac{d[A]}{dt} = pA-nA&\\
&\frac{d[F]}{dt} = pF-nF&\\
&\frac{d[FH]}{dt} = pFH-nFH&\\
&\frac{d[N]}{dt} = pN-nN&
\end{flalign}

\begin{align}
&p0 = 0 &\\
&n0 = k_{upt}\,[Ctr1/3_{trim}] &\\
&p1 = S_{Ctr1/3} + k_{dis1}\,[Ctr1/3_{trim}] &\\
&n1 = k_{b1}\,[Ctr1/3]\,[Ctr1/3]\,[Ctr1/3] + k_{deg1}\,[Ctr_1] + \nonumber\\&k_{Cu-deg}\,[Ctr1/3] &\\
&p2 = k_{b1}\,[Ctr1/3]\,[Ctr1/3]\,[Ctr1/3] + \nonumber\\&k_{tran1}\,[Ctr1/3_{trim}Cu]\,[GSH] \label{trimerization}&\\
&n2 = k_{dis1}\,[Ctr1/3_{trim}] + k_{upt}\,[Ctr1/3_{trim}] &\\
&p3 = k_{upt}\,[Ctr1/3_{trim}] &\\
&n3 = k_{tran1}\,[Ctr1/3_{trim}Cu]\,[GSH] &\\
&p4 = S_{GSH} + k_{tran2}\,[GCu]\,[Ccs1] + k_{tran3}\,[GCu]\,[Atx1] \nonumber\\&+ k_{met}\,[Ace1] &\\
&n4 = k_{deg2}[GSH] + k_{tran1}[GSH][C1_{trim}Cu] &\\
&p5 = k_{tran1}\,[GSH]\,[C1_{trim}Cu] &\\
&n5 = k_{tran2}\,[GCu]\,[Ccs1] + k_{tran3}\,[GCu]\,[Atx1] + \nonumber\\&k_{met}\,[Ace1] &\\
&p6 = S_{Ccs1} + k_{tran5}\,[Ccs1Cu]\,[Sod1] &\\
&n6 = k_{deg3}\,[Ccs1] + k_{tran2}\,[GCu]\,[Ccs1] &\\
&p7 = k_{tran2}\,[GCu]\,[Ccs1] &\\
&n7 = k_{tran5}\,[Ccs1Cu]\,[Sod1] &\\
&p8 = S_{Sod1} &\\
&n8 = k_{deg4}\,[Sod1] + k_{tran5}\,[Ccs1Cu]\,[Sod1] &\\
&p9 = S_{Atx1} + k_{tran6}\,[Atx1Cu]\,[Ccc2] &\\
&n9 = k_{tran3}\,[GCu]\,[Atx1] + k_{deg5}\,[Atx1] &\\
&p10 = k_{tran3}\,[Atx1]\,[GCu] &\\
&n10 = k_{tran6}\,[Atx1Cu]\,[Ccc2] &\\
&p11 = S_{Ccc2} &\\
&n11 = k_{deg6}\,[Ccc2] + k_{tran6}\,[Atx1Cu][Ccc2] &\\
&p12 = S_{Ace1} &\\
&n12 = k_{deg7}\,[Ace1] + k_{met}\,[Ace1]&\\
&p13 = k_{met}\,[Ace1] &\\
&n13 = k_{tran7}\,[Ace1Cu] &\\
&p14 = f_{act} &\\
&n14 = k_{deg,GFPmRNA}\,[GFP_{mRNA}] &\\
&p15 = k_{tl,GFP}\,[GFP_{mRNA}] &\\
&n15 = k_{mat}\,[GFP_{nascent}] &\\
&p16 = k_{mat}\,[GFP_{nascent}] &\\
&n16 = k_{deg,GFPmat}\,[GFP_{mature}]&\\
&p_{luxAB_{mRNA}} = f_{act} &\\
&n_{luxAB_{mRNA}} = k_{deg,luxAB_{mRNA}}\,[luxAB_{mRNA}] \label{LuxmRNA}&\\
&p_{LuxAB} = k_{tl,LuxAB}\,[luxAB_{mRNA}] &\\
&n_{LuxAB} = k_{deg,LuxAB}\,[LuxAB] \label{Luxab deg}&\\
&p_{FRP} = 0 &\\
&n_{FRP} = 0 &\\
&p_{RCOOH} = f_{LuxAB}\,k_4 &\\
&n_{RCOOH} = f_{LuxE} &\\
&p_{LuxERCO} = f_{LuxE} &\\
&n_{LuxERCO} = f_{LuxC,A} &\\
&p_A = f_{LuxC,A} &\\
&n_A = f_{LuxAB}\,k_4 &\\
&p_F = f_{LuxAB}\,k_4 &\\
&n_F = f_{FRP,EX} &\\
&p_{FH} = f_{FRP,EX} &\\
&n_{FH} = f_{LuxAB}\,k_4 &\\
&p_N = k_h\,(f_{FRP,EQ}+f_{LuxC,N}) &\\
&n_N = f_{FRP,EQ}+f_{LuxC,N} &
\end{align}

\begin{flalign}
&k_{upt} = \frac{k_{upt,max}\,[Cu]}{K_{MCu}+[Cu]}\label{uptake rate} &\\
&k_{Cu-deg} = \frac{k_{deg,max}\,[Cu]^2}{K_{Mdeg}^2+[Cu]^2}\\
&k_{met} = \frac{k_{met,max}[GCu]^2}{K_{D,Ace1}^2 + [GCu]^2}\label{metalation rate2} &\\
&f_{act} = \frac{k_{tr,max}}{\sum_{n=1}^{4} \bigg(\frac{K_{D,mRNA}}{[Ace1Cu]}\bigg)^n+1}\label{transcription rate} &\\
&f_{LuxE} = \frac{V^+_{LuxE}\,[RCOOH]\,[ATP]}{D_{LuxE}}\label{LuxE rate}&\\
&f_{LuxC,A} = f_{LuxC,N}= \frac{V^+_{LuxC}\,[LuxE-RCO]\,[N]}{D_{LuxC}}\label{LuxC rate} &\\
&f_{FRP,EX} = f_{FRP,EQ} = \frac{V^+_{FRP}\,[F]\,[N]}{D_{FRP}}\label{Frp rate} &\\ 
&f_{LuxAB} = \frac{k_5\,k6\,k_1\,k_O\,k_2\,k_3\,[FH]\,[A]\,[LuxAB]}{D_{Lux}}\label{Luxab rate} &\\
&D_{FRP} = K_{D,F}\,K_{m,N}+K_{m,N}[F]+K_{m,F}\,[N]+[F]\,[N] \label{Frp denom}&\nonumber\\
& D_{LuxE} = K_{D,RCOOH}\,K_{m,ATP} + K_{m,ATP}\,[RCOOH]+&\nonumber\\&K_{m,RCOOH}\,[ATP]+[RCOOH]\,[ATP] &\nonumber\\
&D_{LuxC}=K_{D,LuxE-RCO}\,K_{m,N2} + K_{m,N2}\,[LuxE-RCO]+\nonumber\\&K_{m,LuxE-RCO}\,[N]+[LuxE-RCO]\,[N] &\nonumber\\
&D_{Lux}=k_2\,k_3\,k_4\,k_5\,k_6\,(k_O+k_{-1})[A] \nonumber\\+ &k_O\,k_1\,k_4\,k_5\,k_6\,(k_3+k_{-2})[FH] \nonumber\\+ &k_1\,k_2(k_3\,k_4\,k_5\,k_6+k_O\,k_4\,k_5\,k_6+k_O\,k_3\,k_5\,k_6+k_O\,k_3\,k_4\,k_6+\nonumber\\&k_O\,k_3\,k_4\,k_5)\,[FH]\,[A] \nonumber
\end{flalign}

\setlength{\tabcolsep}{1pt}

\begin{table}[!htbp]
  \caption{Model Parameters.}\label{params}
  \centering
  \begin{tabular}{|c|p{2.8cm}|p{1.8cm}|p{1.8cm}|}
  \hline
Parameter&Description&Value&Reference\\
\hline
$k_{upt,max}$&Maximum Cu uptake rate&0.0033 $min^{-1}$& approximated \cite{Lin1990}\\
   \hline
$k_{MCu}$&Michaelis Constant for copper uptake&4.4 $\mu$M&\cite{Lin1990}\\
   \hline
$k_{deg,max}$&Maximum Ctrs degradation rate, dependent on Cu&0.01 $min^{-1}$&inferred \cite{Pena2000}\\
   \hline
$k_{M,deg}$&Half-maximum activation constant for copper-dependent degradation of Ctr proteins&20 $\mu M$&inferred \cite{Pena2000}\\
   \hline
$k_{b1}$&Binding rate of Ctr monomers&10 $\mu M^{-2}\,min^{-1}$& assumed\\
\hline
$k_{dis1}$&Unbinding rate of Ctr trimers&1 $min^{-1}$& assumed\\
\hline
$k_{met,max}$&Maximum Ace1 metalation rate due to Cu binding to Ace1&0.1 $min^{-1}$&inferred \cite{furst1989}\\
   \hline
$k_{D,Ace1}$&Half maximum activation constant of Ace1&$0.1$$\mu$M&\cite{furst1989}\\
   \hline
$k_{tr,max}$&Maximum $GFP$ transcription rate from $CUP1$ promoters&2.5 $min^{-1}$&approximated \cite{Mehta2018}\\
   \hline
$k_{D,mRNA}$& Dissociation constant for Ace1 binding to the MREs of the DNA promoter&$10^{-9}$ $\mu$M& assumed\\
   \hline
$k_{tran1}$&Cu handoff rate between $Ctr1/3$ and $GSH$ &0.4 $\mu M^{-1}\,min^{-1}$&approximated \cite{Chacn2018}\\
   \hline
$k_{tran2}$&Cu handoff between $GSH$ and $Ccs1$ &0.4 $\mu M^{-1}\,min^{-1}$&approximated \cite{Chacn2018}\\
   \hline
$k_{tran3}$&Cu handoff rate between $GSH$ and $Atx1$ &0.4 $\mu M^{-1}\,min^{-1}$&approximated \cite{Chacn2018}\\
   \hline
$k_{tran4}$&Cu handoff rate between $GSH$ and $Ace1$ &0.4 $\mu M^{-1}\,min^{-1}$&approximated \cite{Chacn2018}\\
   \hline
$k_{tran5}$&Cu handoff rate between $Ccs1$ and $Sod1$ &0.4 $\mu M^{-1}\,min^{-1}$&approximated \cite{Chacn2018}\\
   \hline
$k_{tran6}$&Cu handoff rate between $Atx1$ and $Ccc2$ &0.4 $\mu M^{-1}\,min^{-1}$&approximated \cite{Chacn2018}\\
   \hline
$k_{tran7}$&Transport rate of metalated Ace1 towards the DNA promoter &0.4 $min^{-1}$&assumed\\
   \hline
$k_{deg1}$& Ctr1/3 monomer degradation rate&0.077 $min^{-1}$& approximated \cite{Pena2000}, \cite{Belle2006}, \cite{Christiano2014}\\
   \hline
$k_{deg2}$& GSH degradation rate&0.0012 $min^{-1}$&approximated \cite{Belle2006}, \cite{Christiano2014}\\
   \hline
$k_{deg3}$& Ccs1 degradation rate&0.0014 $min^{-1}$&\cite{Engel2024}\\
   \hline
$k_{deg4}$& Sod1 degradation rate&0.001 $min^{-1}$&\cite{Engel2024}\\
   \hline
$k_{deg5}$& Atx1 degradation rate&0.0011 $min^{-1}$&\cite{Engel2024}\\
\hline
$k_{deg6}$& Ccc2 degradation rate&0.006 $min^{-1}$&\cite{MartnPrez2015}\\
   \hline
$k_{deg7}$& Ace1 degradation rate&0.04 $min^{-1}$&inferred \cite{Das2017}\\
 \hline 
$k_{deg,GFPmRNA}$& \textit{GFP} mRNA degradation rate&0.12 $min^{-1}$& \cite{Mehta2018}, \cite{Hyde2002}\\
   \hline
   \end{tabular}
\end{table}

\setlength{\tabcolsep}{0.5pt}
\begin{table}[!htbp]
  \caption{Model Parameters (continued).}\label{params2}
  \centering
  \begin{tabular}{|c|p{2.8cm}|p{1.8cm}|p{1.8cm}|}
   \hline     
$k_{tl,GFP}$&Average translation rate in yeast&1.76 $min^{-1}$&\cite{Futcher1999}\\
\hline
$k_{mat}$&GFP maturation rate&0.0058 $min^{-1}$&approximated from \cite{Heim1995}\\
\hline
$k_{deg,GFP_{mat}}$&mature GFP degradation rate&0.0022 $min^{-1}$ assumed based on the long half life of GFP&assumed (long half-life)\\
\hline
$S_{Ctr1/3}$& Constitutive $Ctr1/3$ synthesis rate&0.0001 $\mu M\,min^{-1}$&assumed\\
   \hline
$S_{GSH}$& Constitutive $GSH$ synthesis rate&0.004 $\mu M\,min^{-1}$&assumed\\
   \hline
$S_{Ccs1}$& Constitutive $Ccs1$ synthesis rate&0.0001 $\mu M\,min^{-1}$&assumed\\
   \hline
$S_{Atx1}$& Constitutive $Atx1$ synthesis rate&0.001 $\mu M\,min^{-1}$&assumed\\
   \hline
$S_{Ace1}$& Constitutive $Ace1$ synthesis rate&0.0003 $\mu M\,min^{-1}$&assumed\\
   \hline
$S_{Sod1}$& Constitutive $Sod1$ synthesis rate&0.0001 $\mu M\,min^{-1}$&assumed\\
   \hline
$S_{Ccc2}$& Constitutive $Ccc2$ synthesis rate&0.00001 $\mu M\,min^{-1}$&assumed\\
   \hline
$k_{deg,luxAB_{mRNA}}$& \textit{luxAB} mRNA degradation rate&0.09 $min^{-1}$&assumed\\
   \hline
$k_{tl,LuxAB}$&average translation rate in yeast&1.76 $min^{-1}$&\cite{Futcher1999}\\
\hline
$k_{deg,LuxAB}$& degradation rate of LuxAB&0.003 $min^{-1}$& \cite{Tinikul2012}\\
\hline
$V^+_{LuxE}$& forward catalysis rate for LuxE&1 $\mu M\,min^{-1}$& approximated \cite{Welham2008}\\
\hline 
$V^+_{LuxC}$& forward catalysis rate for LuxC&1 $\mu M\,min^{-1}$&\cite{Welham2008}\\
\hline
$ K_{D,RCOOH}$& Dissociation constant of fatty acids&0.01 $\mu$M&assumed\\
\hline
[ATP]& ATP concentration& 2500 $\mu$M& \cite{zalp2010}\\
\hline
$ K_{m,ATP}$& Michaelis constant for ATP& 0.002 $\mu$M& \cite{Welham2008}\\
\hline
$K_{m,RCOOH}$& Michaelis constant for fatty acids&0.04342 $\mu$M&\cite{Welham2008}\\
\hline
$K_{m,LuxE-RCO}$& Michaelis constant for the acyl-LuxE intermediate complex&0.2 $\mu$M& assumed\\
\hline
$K_{D,LuxE-RCO}$&Dissociation constant for the acyl-LuxE intermediate complex&0.001 $\mu$M&assumed\\
\hline
$K_{m,N2}$& Michaelis constant for NADPH (LuxC reaction)&5 $\mu$M&\cite{Welham2008}\\
\hline
$K_{m,N}$& Michaelis constant for NADPH (FRP reaction)&9 $\mu$M&\cite{Chung2012}\\
\hline
$K_{m,F}$& Michaelis constant for FMN&0.1117 $\mu$M&calculated \cite{Musila2018}, \cite{cornish2013}\\
\hline
$K_{D,F}$& Dissociation constant of FMN&0.017 $\mu$M&\cite{Musila2018}\\
\hline
$k_{cat}$& Constant rate for FRP catalysis&260 $min^{-1}$& calculated \cite{Musila2018}, \cite{cornish2013}\\
\hline
$k_h$& Constant rate for re-formation of NADP to $NADPH$ after recycling& 1&assumed\\
\hline
$K_E$& equilibrium constant for protonation-deprotonation&$1.6\cdot 10^{-8}$ $\mu$M&\cite{Tinikul2020B}\\
\hline
[h]& hydrogen concentration& 0.16 $\mu$M& \cite{Milo2009}, BNID 107762\\
\hline
$k_O$&reaction rate constant of oxygen&21000 $min^{-1}$&\cite{TINIKUL2020}\\
\hline
$k_1$&binding rate of $\mathrm{FMNH_2}$&1020 $\mu M^{-1}\,min^{-1}$&\cite{AbuSoud1992}\\
\hline
   \end{tabular}
\end{table}

\setlength{\tabcolsep}{0.5pt}
\begin{table}[!htbp]
  \caption{Model Parameters (continued).}\label{params3}
  \centering
  \begin{tabular}{|c|p{3cm}|p{2.5cm}|p{2cm}|}
  \hline
$k_{-1}$&unbinding rate of $\mathrm{FMNH_2}$& 7200 $min^{-1}$&\cite{AbuSoud1992}\\
\hline
$k_{bA}$&binding rate of aldehyde& 600 $\mu M^{-1}\,min^{-1}$&\cite{AbuSoud1992}\\
\hline
$k_{-2}$& Unbinding rate for aldehyde&18000 $min^{-1}$&\cite{AbuSoud1992}\\
\hline
$k_3$& hydroxyflavin formation rate&15 $min^{-1}$&approximated \cite{TINIKUL2020}\\
\hline
$k_4$& hydroxyflavin de-excitation and de-hydration rate&0.0001 $min^{-1}$& approximated \cite{TINIKUL2020}, \cite{AbuSoud1992}\\
\hline
$k_5$& fatty acid RCOOH release rate&46 $min^{-1}$& \cite{AbuSoud1992}\\
\hline
$k_6$& $FMN$ release rate&36 $min^{-1}$&inferred \cite{AbuSoud1992}\\
\hline
$k_L$&photon emission rate& 144 $min^{-1}$ &\cite{Tinikul2012}\\
\hline
$R_{pd}$& photodiode responsivity&0.185 $A/W$ (at $\lambda=490$ nm)&\cite{Koklu2013}\\
\hline
   \end{tabular}
\end{table}

\setlength{\tabcolsep}{1pt}
\begin{table}[!htbp]
  \caption{Initial values of the various chemical compounds within the Copper Uptake Pathway.}
  \label{initial value}
  \centering
  \begin{tabular}{|c|c|p{2cm}|}
  \hline
Chemical Compound&Initial Concentration ($\mu$M)&Reference\\
  \hline
Ctr1/3&0.1 &\cite{Engel2024}\\
   \hline
$\mathrm{Ctr1/3_{trim}}$&0.03& assumed\\
\hline
GSH&300&\cite{Tahmasebi2016}\\
   \hline
Ccs1&0.25&approximated \cite{Engel2024}\\
   \hline
Sod1&2.2&approximated \cite{Engel2024}\\
   \hline
Atx1&0.22&approximated \cite{Engel2024}\\
   \hline
Ccc2&0.034&approximated \cite{Engel2024}\\
   \hline
Ace1&0.0428& approximated \cite{Engel2024}\\
   \hline
FRP&1000&assumed\\
\hline
RCOOH&8&assumed\\
   \hline
F&500& saturating concentration \cite{Chung2012}, \cite{Musila2018}\\
\hline
N&500& saturating concentration \cite{Chung2012}, \cite{Musila2018}\\
\hline
\end{tabular}
\end{table}

\newpage
\section*{Acknowledgments}
This research has received support from the projects BRIDGE2HORIZON/0823D/004 (YeasPSense) and EXCELLENCE/0524/0595 (YeCIE) which is implemented under the Cohesion Policy Funds “THALIA 2021-2027” with EU co-funding.



\bibliography{Refs}{}
\bibliographystyle{IEEEtran}

\vfill

\end{document}